\documentclass[10pt,journal,compsoc]{IEEEtran}

\usepackage{graphicx,amsmath,amssymb,booktabs,multirow,array,subfigure,url,paralist,amsmath,color}
\usepackage{bm,amsmath}
\usepackage[ruled,linesnumbered]{algorithm2e}
\usepackage[hidelinks]{hyperref}

\newcommand\impath{./figures}

\DeclareMathOperator*{\argmin}{argmin}

\begin{document}

\title{Explore the Effect of Data Selection on Poison Efficiency in Backdoor Attacks}

\author{
	Ziqiang Li, Pengfei Xia, Hong Sun, Yueqi Zeng, Wei Zhang, and Bin Li, \IEEEmembership{Member, IEEE}
	\IEEEcompsocitemizethanks{
		\IEEEcompsocthanksitem Z. Li, P. Xia, H. Sun, Y. Zeng, W. Zhang, and B. Li are with the Department of Electronic Engineering and Information Science, University of Science and Technology of China, Hefei, China. \\ E-mail: \{xpengfei, iceli, hsun777, zyueqi, zw1996\}@mail.ustc.edu.cn, binli@ustc.edu.cn.
		\IEEEcompsocthanksitem An earlier version of this study has been presented in IJCAI 2022.
	}
        \thanks{Equal Contribution: Ziqiang Li, Pengfei Xia, Hong Sun.}
	\thanks{Corresponding author: Bin Li.}
}

\IEEEtitleabstractindextext{
\begin{abstract}

As the number of parameters in Deep Neural Networks (DNNs) scales, the thirst for training data also increases. To save costs, it has become common for users and enterprises to delegate time-consuming data collection to third parties. Unfortunately, recent research has shown that this practice raises the risk of DNNs being exposed to backdoor attacks. Specifically, an attacker can maliciously control the behavior of a trained model by poisoning a small portion of the training data. In this study, we focus on improving the poisoning efficiency of backdoor attacks from the sample selection perspective. The existing attack methods construct such poisoned samples by \textit{randomly} selecting some clean data from the benign set and then embedding a trigger into them. However, this random selection strategy ignores that each sample may contribute differently to the backdoor injection, thereby reducing the poisoning efficiency. To address the above problem, a new selection strategy named Improved Filtering and Updating Strategy (FUS++) is proposed. Specifically, we adopt the forgetting events of the samples to indicate the contribution of different poisoned samples and use the curvature of the loss surface to analyses the effectiveness of this phenomenon. Accordingly, we combine forgetting events and curvature of different samples to conduct a simple yet efficient sample selection strategy. The experimental results on image classification (CIFAR-10, CIFAR-100, ImageNet-10), text classification (AG News), audio classification (ESC-50), and age regression (Facial Age) consistently demonstrate the effectiveness of the proposed strategy: the attack performance using FUS++ is significantly higher than that using random selection for the same poisoning ratio.
\end{abstract}

\begin{IEEEkeywords}
Deep Neural Networks, Backdoor Attacks, Poisoning Efficiency, Sample Selection, Curvature.
\end{IEEEkeywords}
}
\maketitle

\section{Introduction} 
\label{sec:int}
Despite the remarkable accomplishments demonstrated by Deep Neural Networks (DNNs) across various tasks \cite{silver2016mastering, krizhevsky2017imagenet, yin2019online, xia2020boosting, jumper2021highly}, their susceptibility to malicious attacks, such as adversarial examples \cite{szegedy2013intriguing, goodfellow2014explaining, xia2021improving, xia2021tightening}, remains a critical concern. Particularly during the training phase, a significant security challenge arises in the form of \textit{backdoor attacks} or \textit{Trojan attacks} \cite{chen2017targeted, gu2017badnets, xia2022enhancing}. These attacks involve the insertion of a small number of tainted samples into the legitimate training dataset, thereby enabling the trained DNN to be covertly implanted with a concealed backdoor. Following training, the compromised model exhibits normal behavior when presented with benign inputs, rendering the attack detection process arduous. However, upon activation of a predefined trigger, the backdoor is invoked, causing the model's predictions to align with the objectives of the attacker. Notably, the escalating demand for diverse training data for DNNs \cite{brown2020language, radford2021learning} has led to the common practice of sourcing data from the internet and other unfamiliar sources, thereby creating a feasible avenue for the execution of backdoor attacks.


An emerging trend within the realm of backdoor attack development aims to enhance stealthiness to effectively circumvent detection by both human analysts and algorithmic systems. To exemplify, Zhong \textit{et al.} \cite{zhong2020backdoor} introduced the concept of employing imperceptible noise as the triggering mechanism, a departure from the previously employed conspicuous patterns \cite{gu2017badnets} or images \cite{chen2017targeted}, thereby ensuring minimal visibility to observers. Turner \textit{et al.} \cite{turner2019label} emphasized the potential suspicion raised by the incongruity between the semantic essence of a poisoned sample and its designated label. To mitigate this concern, they advanced the notion of exclusively incorporating the trigger into the clean data affiliated with the targeted class, while concurrently augmenting the potency of poisoning through techniques such as adversarial examples \cite{madry2017towards} and generative adversarial networks \cite{goodfellow2020generative}.

\begin{figure*}[t]
\begin{center}
\includegraphics[width=0.9\textwidth]{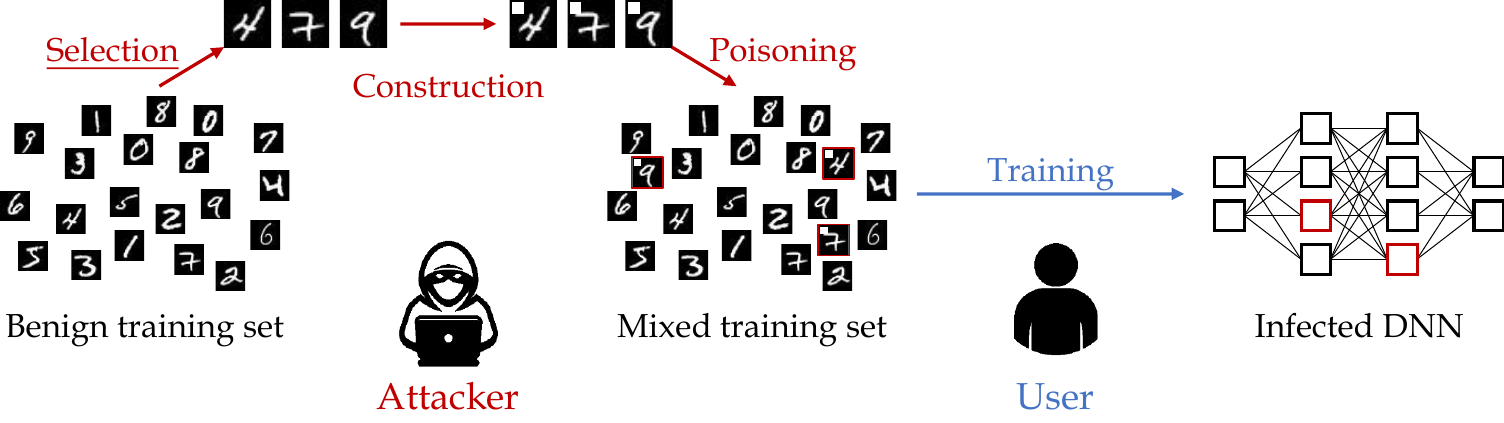}
\end{center}
\caption{The brief flow of poisoning-based backdoor attacks. The attacker uses the selection, construction, and poisoning steps to build a mixed training set and releases it. The user gets this training set to train a (backdoored) DNN. For the attacker, the number of poisoned samples in the mixed training set may affect the stealthiness of the attack. This study focuses on the \underline{selection} step to improve the poisoning efficiency, thereby improving the poisoned stealthy.}
\label{fig:att_flo}
\end{figure*}

Beyond the aforementioned considerations, the quantity of poisoned samples significantly influences the surreptitious nature of backdoor attacks. Generally, a higher count of poisoned samples increases the likelihood of successfully embedding a backdoor, but concurrently elevates the risk of exposure. Hence, it becomes imperative to explore strategies for diminishing the volume of poisoned samples while retaining the potency of the attack. To accomplish this objective, we begin by revisiting the progression of poisoning-based backdoor attacks, illustrated in \figurename~\ref{fig:att_flo}. The attacker engenders the amalgamated training set through three sequential stages:
\begin{inparaenum}
\item \textit{Selection:} Extraction of clean data from the benign training set;
\item \textit{Construction:} Utilization of the selected data to formulate poisoned samples;
\item \textit{Poisoning:} Reintegration of the fabricated poisoned samples into the benign training set.
\end{inparaenum}
Existing endeavors \cite{zhao2020clean, zhong2020backdoor} that strive to enhance the efficiency of the poisoning process have primarily concentrated on the construction phase. For instance, Zhao \textit{et al.} \cite{zhao2020clean} advocated for the adoption of an optimized trigger in lieu of a static one, illustrating that the former yields a heightened attack success rate in the context of video recognition tasks \cite{yue2015beyond}, even when maintaining the same poisoning ratio.

Nevertheless, scant attention has been directed towards the preliminary selection step, specifically the crucial question of \textbf{identifying the appropriate benign samples for poisoning}. The majority of extant attack techniques \cite{chen2017targeted, gu2017badnets, turner2019label, nguyen2021wanet} resort to a \textit{random} selection process from the benign training set. This indiscriminate approach implicitly assumes that each chosen sample contributes equally to the backdoor implantation—an assumption that proves challenging to uphold in practical scenarios. Consequently, this assumption raises a pertinent issue: the attack's efficiency may suffer due to the presence of numerous low-contributing poisoned samples within the amalgamated training dataset. In this context, our study addresses this overlooked concern and introduces an innovative selection strategy to address it. Specifically, we formulate a discrete constrained optimization problem to facilitate effective sample selection for poisoned attacks. To navigate the intricacies of this optimization problem, we propose a pragmatic iterative selection strategy aimed at approximating solutions. At the heart of our approach lies the fundamental principle of identifying those poisoned samples that exert a more substantial impact on the backdoor injection, utilizing them to construct the poisoned dataset. Drawing inspiration from prior research \cite{katharopoulos2018not, toneva2018empirical} within the realm of deep learning, we delve into the correlation between forgettable instances and sample efficiency. Our analysis reveals that samples prone to being forgotten during the poisoning process assume a pivotal role in constructing potent poison attacks. Furthermore, to elucidate this phenomenon, we delve into recent investigations concerning the curvature of the loss surface \cite{garg2023samples, dinh2017sharp}. This exploration affords insights into the sample efficiency of selectively chosen samples through the lens of forgettable events, underscoring the correlation between selected efficient samples and low-curvature attributes. In light of these revelations, we introduce an elegantly straightforward yet remarkably effective sample selection strategy: the Improved Filtering and Updating Strategy (FUS++). This strategy notably enhances poisoned efficiency in comparison to conventional selection methodologies. To summarize, our contributions encompass four pivotal dimensions:

\begin{itemize}

\item To the best of our knowledge, the proposed strategy is the first method to improve the poisoning efficiency of backdoor attacks from the sample selection perspective, which remains orthogonal to prior investigations.
\item Our study stands as the inaugural exploration into quantifying sample efficiency within poisoning-based backdoor attacks, accomplished through a meticulous analysis of both forgetting events and the curvature of the loss surface for each sample. This groundbreaking approach furnishes a fresh and insightful understanding of sample selection within the context of backdoor attacks.
\item We formulate a discrete constrained optimization problem to optimize the selection of samples and thereby enhance poisoning efficiency—a challenge marked by its intricacy. Employing systematic and comprehensive analyses, we introduce a strategy named the Improved Filtering and Updating Strategy (FUS++) that provides a practical solution to this complex problem.
\item Our proposed strategy is subjected to rigorous evaluation across four diverse tasks: image classification, text classification, audio classification, and age regression. The results consistently affirm its efficacy. For instance, on CIFAR-10 with an identical 0.01 mixing ratio, the attack success rate utilizing FUS++ is 0.923, as opposed to 0.871 with random selection.

\end{itemize}

An earlier version of this study has been presented in IJCAI 2022\footnote{The code for the earlier conference version of FUS is accessible through the following link: \hyperref[Github]{https://github.com/xpf/Data-Efficient-Backdoor-Attacks}. Furthermore, we intend to make the code for our enhanced FUS++, which is the focus of this work, available on the same platform.} \cite{xia2022data}. This journal manuscript represents a substantial expansion upon its initial conference version in terms of methodological enhancements, comprehensive experimentation, and refined exposition. The details of these improvements are outlined below:

1. We have significantly enhanced the previous FUS method across various dimensions:
\begin{itemize}
\item The conference version of FUS primarily relied on forgetting events as a measure of sample importance, which was time-intensive and tailored to classification tasks exclusively. In contrast, this journal version introduces a more generalized definition (Sec. \ref{sec: fus}). We employ three key indicators to gauge the importance of poisoned samples: forgetting events, confidence scores (for classification tasks), and loss swings (for regression tasks).
\item Unlike the conference version's sole use of the Fixed Filtration Ratio policy, we extend our approach in this journal version to encompass multiple filtration ratio policies, including Fixed, Linear Decay, and Exponential Decay.
\item Recognizing the substantial number of iterations required to complete the search in FUS and the consequent linear increase in the attacker's computational expenses, particularly in terms of retraining the model over multiple epochs in each iteration, it becomes evident that employing this strategy in the context of large or complex models is rendered impractical. To save the time cost of the attacker, our enhanced FUS++ (Sec. \ref{sec:Curvature}, \ref{sec:low_curvature}, and \ref{sec:fus_plus}) incorporates the curvature of the loss surface in the vicinity of samples to identify high-contributing samples. This enhancement, which involves a straightforward method for obtaining the curvature of the loss surface, filters out low-contributing samples effectively during the poisoning process with a very low cost. By subsequently applying FUS to the subset of high-contributing samples selected based on curvature, we only need one-step iteration in FUS to acquire the best poisoning efficiency (Figure. \ref{fig:cla_N}).
\end{itemize}
2.  We offer systematic and exhaustive analyses to elucidate the reasons underlying the superior performance of FUS-selected poisoned samples compared to random samples (Sec. \ref{sec:att_studies}):

\begin{itemize}
\item We investigate the class distribution of FUS-selected poisoned samples, demonstrating that the class distribution acts as a symptom rather than the cause of sample efficiency.
\item Our exploration delves into the influence of different triggers, revealing that the high efficiency of FUS-selected samples results from the interaction between the sample and the trigger, rather than an inherent property of the sample itself.
\item In addition, we probe the resilience of various methods against backdoor defense, demonstrating that FUS exhibits superior performance in countering backdoor defenses compared to the random selection strategy.
\end{itemize}
3. The experimental segment has been significantly enriched:
\begin{itemize}
\item Compared to FUS, our proposed FUS++ further improves the poisoning efficiency (Sec. \ref{sec:results_image_classi}, Sec. \ref{sec:results_text_classi}, Sec. \ref{sec:results_audio_class}, Sec. \ref{sec:results_age_reg}, and Sec. \ref{sec:results_limited}), and achieves the state of the art (Sec. \ref{sec:results_sota}).
\item While the conference version evaluated FUS solely in the context of image classification with flip-label attacks, this journal version broadens the scope by including three additional tasks: text classification (Sec. \ref{sec:results_text_classi}), audio classification (Sec. \ref{sec:results_audio_class}), and age regression (Sec. \ref{sec:results_age_reg}), encompassing diverse modalities to test FUS++ across different domains.
\item We undertake a comprehensive array of ablation studies, specifically focusing on dissecting the distinct components constituting our FUS++ methodology, as expounded in Sec. \ref{sec:ablation}.
\item We add the comparison with previous state-of-the-art methods of sample selection on poisoned attacks (Sec. \ref{sec:results_sota}).
\item Our evaluation is extended to encompass various attack settings, including clean-label attacks (Sec. \ref{sec:results_image_classi}) and limited data attacks (Sec. \ref{sec:results_limited}).
\end{itemize}

\section{Related Work}
\subsection{Backdoor Attacks}

Backdoor attacks are devised to clandestinely insert concealed Trojans into Deep Neural Networks (DNNs), thereby exerting control over their behavior. This threat landscape was initially explored by Gu \textit{et al.} \cite{gu2017badnets}, and since its inception, numerous variations have emerged. Broadly categorized based on their methods of backdoor injection, these attack techniques fall into two distinct groups \cite{li2022backdoor}: poisoning-based attack methods \cite{liu2017trojaning, nguyen2021wanet} and non-poisoning-based attack methods \cite{dumford2018backdooring, kurita2020weight}. The nomenclature succinctly captures their essence—the former group orchestrates Trojan implantation by manipulating the integrity of the clean training dataset, while the latter group employs mechanisms such as transfer learning or direct modification of DNN weights to execute attacks. In this investigation, we narrow our focus to the first category of backdoor attacks.


The prevailing body of research \cite{turner2019label, li2020invisible, zhao2020clean, zhong2020backdoor} pertaining to poisoning-based attack methods has primarily concentrated on strategies to enhance stealthiness, thereby evading detection by both human observers and algorithmic systems. To illustrate, Zhong \textit{et al.} \cite{zhong2020backdoor} recognized that commonly employed triggers, such as local patterns \cite{gu2017badnets} or illustrative images \cite{chen2017targeted}, possess a visual presence perceptible to human eyes. In response, they advocated for the utilization of additive noise as a trigger while ensuring its visual invisibility by constraining its amplitude. An analogous objective was achieved by Hammoud and Ghanem \cite{hammoud2021check}, who opted to embed triggers within the frequency domain of images rather than their spatial domain. In addition to visual visibility, another aspect under scrutiny is the discordance between the content of an image and its assigned label \cite{turner2019label, zhao2020clean}. When constructing a poisoned sample, the inclusion of a trigger within a clean sample is often requisite, accompanied by the setting of its label to reflect the attack target. Turner \textit{et al.} \cite{turner2019label} brought attention to this quandary and proposed clean-label backdoor attacks as a countermeasure. This approach effectively addresses the inconsistency between image content and labels in the context of poisoning-based attacks.

This study centers its attention on the quantity of poisoned samples integrated into the training dataset, a factor that also substantially impacts the stealthiness of the attack. In contrast to preceding research, we offer a distinct vantage point: enhancing the efficiency of backdoor attacks through sample selection. Notably, our conference version \cite{xia2022data} is among the pioneering endeavors to delve into the realm of sample selection within backdoor attacks. Additionally, subsequent investigations \cite{gao2023not, li2023proxy} have pursued this line of inquiry. Among these, Gao \textit{et al.} \cite{gao2023not} introduce three classical difficulty metrics—Loss Value, Gradient Norm, and Forgetting Event—to quantify challenging samples during clean-label backdoor attacks, with the Forgetting Event emerging as the most effective selection strategy. It is noteworthy that their approach diverges from our FUS and FUS++ in adopting a greedy selection strategy, constituting a specific instance of our iterative FUS approach. The mechanics of the greedy selection strategy are detailed in Sec. \ref{sec:Greedy_selection}. Simultaneously, Li \textit{et al.} \cite{li2023proxy} scrutinize the limitations of FUS when discrepancies arise in data transformation and optimization hyperparameters between proxy and actual attacks. Consequently, they propose a proxy-free selection strategy (PFS) that achieves comparable or even superior poisoned efficiency to FUS via an elegantly straightforward approach. PFS rests on the empirical observation that utilizing samples exhibiting high similarity between clean and corresponding poisoned variants yields significantly higher attack success rates compared to the use of low-similarity samples. In a departure from this, we introduce the concept of the curvature of the loss surface to offer a theoretical rationale for varying sample contributions during backdoor attacks. Our newly proposed FUS++, building on these insights, surpasses the performance of other selection methods.

\subsection{Data-efficient Learning}

Deep learning models have a voracious appetite for data, yet collecting and annotating data is often a time-consuming and costly endeavor, particularly in specialized sectors like aviation and aerospace. This scenario has spurred efforts to enhance data utilization efficiency, a pursuit known as data-efficient learning \cite{popov2017data, adadi2021survey, li2022comprehensive}. Categorized according to \cite{adadi2021survey}, prevailing data-efficient techniques fall into four domains:
\begin{itemize}
\item Non-supervised learning: In domains where acquiring substantial data is feasible but labeling proves challenging, non-supervised paradigms have emerged as a solution. These paradigms aid Deep Neural Networks (DNNs) in acquiring effective representations. Examples include contrastive learning \cite{chen2020simple, he2020momentum} and masked modeling \cite{he2022masked}.

\item Data augmentation: This approach bolsters both data quantity and quality and has proven immensely effective \cite{shorten2019survey}. Standard techniques encompass color and geometric transformations \cite{krizhevsky2017imagenet}, image mixing \cite{zhang2017mixup}, and more.

\item Transfer learning: Grounded in the concept that knowledge garnered from data-rich domains can be transferred to enhance learning in data-scarce domains, transfer learning \cite{long2017deep, zhuang2020comprehensive} has demonstrated efficacy in mitigating data requirements.

\item Data-efficient algorithms: Certain methodologies ensure effective learning from limited data by designing enhanced learning algorithms. Neural-symbolic computing \cite{garcez2019neural} is a notable example.
\end{itemize}

\subsection{Importance Sampling}

The assessment of sample importance is a pivotal and intriguing subject across various domains. Importance sampling \cite{johnson2018training, toneva2018empirical} emerges as a technique that prioritizes essential samples, thereby expediting and enhancing the efficiency of training. Within this context, we narrow our focus to coreset selection, which serves as the foundational method for determining sample importance scores. Drawing inspiration from the insights of \cite{garg2023samples}, existing importance sampling methods can be categorized as follows:
\begin{itemize}
\item Clustering-based methods: By aggregating samples in proximity, clustering-based techniques exploit the notion that samples closely situated can be aptly represented by the cluster center. Consequently, adopting cluster centers \cite{chen2012super, har2004coresets} as representative samples constitutes a straightforward importance sampling approach.

\item Classification output-based methods: These strategies \cite{toneva2018empirical, coleman2019selection} operate under the assumption that samples yielding lower-confidence classifications harbor greater complexity and information compared to instances classified with heightened confidence. Among these, \cite{toneva2018empirical} demonstrates that a substantial proportion of samples are accurately learned early in training and remain consistently retained throughout. These perpetually retained samples, being easily learned, should be pruned from the coreset.

\item Loss gradient-based methods: These methodologies \cite{killamsetty2021grad, mirzasoleiman2020coresets, garg2023samples} establish the importance of a sample by evaluating its contribution to the loss function, its loss gradient, or the Hessian of the loss function. Notably, \cite{garg2023samples} selects coresets from pre-trained models based on the trace of the Hessian of the loss and unveils that the curvature of the loss function corresponds to the concept of visual "cleanliness". Samples with low curvature are devoid of conflicting features, whereas high curvature samples appear intricate, cluttered, or misrepresentative in comparison to other class samples.
\end{itemize}

\section{Methodology}

\begin{figure*}[htbp]
\begin{center}
\subfigure[]{\includegraphics[width=0.32\textwidth]{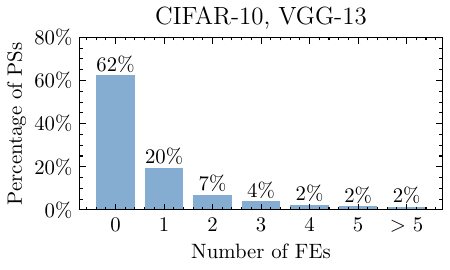} \label{fig:obs_a}}
\subfigure[]{\includegraphics[width=0.32\textwidth]{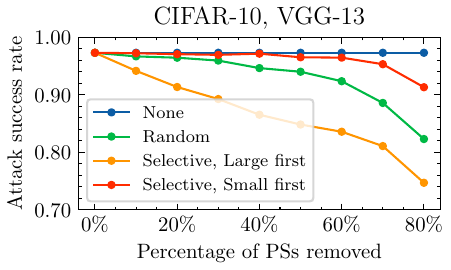} \label{fig:obs_b}}
\subfigure[]{\includegraphics[width=0.32\textwidth]{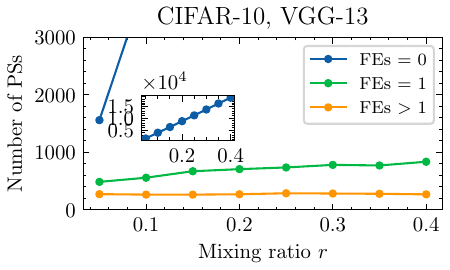} \label{fig:obs_c}} 
\end{center}
\caption{The experimental outcomes pertaining to the forgetting events of poisoned samples on CIFAR-10 and VGG-13 are presented herein, with all values representing averages derived from 5 independent runs. Abbreviations utilized: FEs (Forgetting Events), PSs (Poisoned Samples). (a) Depiction of a histogram illustrating the distribution of forgetting events when $r = 0.05$, wherein approximately 38\% of the poisoned samples experience at least one instance of forgetting during the injection process. (b) Analysis of the attack success rate in relation to the gradual removal of poisoned samples. In comparison to random removal, the utilization of selective removal predicated on ascending forgetting event orders yields superior preservation of attack potency. Conversely, the poorest performance is observed when employing selective removal based on descending forgetting event orders. (c) Examination of the number of poisoning samples across varying forgetting event occurrences as the mixing ratio undergoes modification. Notably, the sample volume for forgetting events surpassing 1 exhibits marginal growth as the ratio escalates. The subplot further highlights the poisoned sample volume for forgetting events equal to 0, displaying linear augmentation in tandem with an increase in the mixing ratio.}
\end{figure*}

\subsection{Problem Formulation}
We specifically concentrate on backdoor attacks that exploit poisoning techniques, as outlined in previous works \cite{chen2017targeted,gu2017badnets}. The attack methodology is illustrated in the flow depicted in \figurename~\ref{fig:att_flo}. To formally define this process, we begin with a benign training set denoted as $\mathcal{D} = {(x, y)}$, alongside a set of poisoned samples referred to as $\mathcal{U} = {(x', t)}$. The procedure involves the introduction of data poisoning by integrating the poisoned set $\mathcal{U}$ into the benign set $\mathcal{D}$, culminating in the creation of a combined training set denoted as $\mathcal{M} = \mathcal{D} \cup \mathcal{U}$. Within this context, $(x, y)$ represents a benign sample along with its corresponding true label, while $(x', t)$ corresponds to a poisoned sample along with the target label determined by the attacker. Subsequently, the attacker presents the manipulated training set $\mathcal{M}$ to the user, who unwittingly treats it as a standard training set and proceeds to train a Deep Neural Network (DNN) using it:
\begin{equation}
\min_{\theta} \ \  \frac{1}{|\mathcal{M}|}\sum_{(\hat{x}, \hat{y}) \in \mathcal{M}} L(f_{\theta}(\hat{x}), \hat{y}),
\end{equation}
where the notation $f_{\theta}$ represents the DNN model and its associated parameters, while $L$ signifies the loss function employed for training. The data pair $(\hat{x}, \hat{y})$ is drawn from the amalgamated dataset $\mathcal{M}$. The primary objective is to train a model that is subtly infected with a concealed backdoor, capable of performing well on both the untainted test set $\mathcal{T}$ and the poisoned test set $\mathcal{V}$. The mixing ratio, denoted as $r$, is defined as the proportion of poisoned samples in relation to the clean samples within $\mathcal{M}$, expressed as $r = |\mathcal{U}| / |\mathcal{D}|$. In broad terms, a higher value of $r$ indicates a more potent backdoor attack, whereas a lower value of $r$ suggests a more covert attack. The goal of this study is to diminish the value of $r$ without compromising the effectiveness of the attack itself.

Illustrated in \figurename~\ref{fig:att_flo}, the process of selecting and assembling the poisoned training set $\mathcal{U}$ assumes paramount importance in determining the efficacy of the attack. When presented with clean data and its corresponding label $(x, y)$ obtained from $\mathcal{D}$, the generation of a corresponding poisoned pair $(x', t)$ becomes possible—wherein $x' = F(x, k)$. In this context, the function $F$ integrates the trigger $k$ into the original input $x$. For instance, a notable approach by Chen \textit{et al.} \cite{chen2017targeted} introduced an image blending attack, generating a poisoned sample through $x' = \lambda \cdot k + (1 - \lambda) \cdot x$, with $\lambda$ representing the blending ratio, and $k$ constituting an image of the same dimensions as $x$. Given that each clean pair within $\mathcal{D}$ has the potential to yield a corresponding poisoned pair, a comprehensive set denoted as $\mathcal{D}' = {(F(x, k), t) | (x, y) \in \mathcal{D}}$ can be established, encapsulating all feasible candidates. The actual formation of $\mathcal{U}$ involves the selection of $r \cdot |D|$ samples from $\mathcal{D}'$, effectively denoted as $\mathcal{U} \subset \mathcal{D}'$. In practical scenarios, the attacker forges $\mathcal{U}$ by cherry-picking certain uncontaminated data instances from $\mathcal{D}$ and subsequently implanting the trigger within them—although the explicit construction of $\mathcal{D}'$ is not a requisite step. The articulation of $\mathcal{D}'$ in this context serves the purpose of providing a clearer depiction of this process.

Two key factors influencing the efficiency of the poisoning process encompass both the construction of pairs $(F(x, k), t)$ and the subsequent selection of $\mathcal{U}$ from $\mathcal{D}'$. While past research \cite{zhong2020backdoor,zhao2020clean} has primarily delved into refining the construction phase, the current study pivots its focus towards the selection stage—a hitherto underexplored facet. Presently, the majority of attack methodologies rely on a randomized selection approach, which disregards the varying significance of individual poisoned samples. Our pursuit centers on enhancing poisoning efficiency through the judicious selection of $\mathcal{U}$ from the pool of $\mathcal{D}'$. This objective can be framed as follows:
\begin{equation}
\label{equ:formulate}
\begin{split}
\max_{\mathcal{U} \subset \mathcal{D}'} \ \  & \frac{1}{|\mathcal{V}|} \sum_{(x', t) \in \mathcal{V}} \mathbb{I}(f_{\theta^*}(x') = t) \\
\text{s.t.} \ \  & \theta^* = \argmin_{\theta} \frac{1}{|\mathcal{M}|} \sum_{(\hat{x}, \hat{y}) \in \mathcal{M}} L(f_{\theta}(\hat{x}), \hat{y})) \\
& \frac{1}{|\mathcal{T}|} \sum_{(x, y) \in \mathcal{T}} \mathbb{I}(f_{\theta^*}(x) = y) \approx \epsilon \\
& |\mathcal{U}| = r \cdot |\mathcal{D}| \\
& \mathcal{M} = \mathcal{D} \cup \mathcal{U} \\
\end{split} \text{,}
\end{equation}
where $\mathbb{I}$ represents the indicator function, and $\epsilon$ assumes the role of a value ensuring that the accuracy of the model trained on $\mathcal{M}$ aligns closely with that of the model trained solely on $\mathcal{D}$. This equation gives rise to a discrete constrained optimization quandary, presenting a formidable challenge to resolve. In response, we put forth a pragmatic, iterative selection strategy aimed at identifying approximate solutions.

\subsection{Low- and High-contributing Samples}
The fundamental premise of our methodology resides in the identification of those tainted samples that yield substantial contributions to the introduction of backdoor effects, subsequently harnessing them for the constitution of $\mathcal{U}$. A preliminary inquiry necessitates clarification: do low- and high-contributing contaminated samples indeed exist within the context of backdoor attacks? To address this query, we direct our attention towards the realm of conventional classification tasks. The quintessential embodiment of sample significance can be found within support vector machines \cite{cortes1995support}, wherein solely the support vectors hold sway over the determination of an optimal hyperplane amidst a voluminous dataset. Within the purview of the deep learning era, various investigations \cite{katharopoulos2018not,toneva2018empirical} have illuminated the notion that certain arduous or easily forgettable samples carry heightened importance in the process of shaping the representation and decision boundary of deep neural networks (DNNs). Given the apparent parity between training a infected model and training a standard model once the amalgamation of poisoned samples has been executed—depicted in \figurename~\ref{fig:att_flo}—we posit the presence of challenging or forgettable poisoned samples. These samples are postulated to assume a pivotal role in influencing the potency of the attack.

To substantiate the aforementioned perspective, we employ the concept of "forgetting events" \cite{toneva2018empirical} to elucidate the learning dynamics associated with each poisoned sample throughout the backdoor injection process. In the work presented by Toneva \textit{et al.} \cite{toneva2018empirical}, an approach is introduced that quantifies the occurrence of forgetting for individual samples over the course of training, ultimately assigning an importance score to each. In our study, we exclusively monitor the occurrence of forgetting events within the realm of poisoned samples encompassed by $\mathcal{M}$. Concretely, consider a poisoned sample along with its target $(x', t)$ drawn from $\mathcal{M}$. If, at time step $s$, $x'$ is classified as the target $t$, i.e., $\mathbb{I}(f_{\theta^s}(x') = t) = 1$, but is subsequently misclassified at time step $s + 1$, i.e., $\mathbb{I}(f_{\theta^{s + 1}}(x') = t) = 0$, this occurrence is logged as a "forgetting event" for that specific sample. Here, $\theta^s$ and $\theta^{s + 1}$ denote the model parameters at moments $s$ and $s + 1$, respectively. As a given poisoned sample may undergo multiple such transitions during the Trojan implantation process, a cumulative tally of forgetting events is recorded and leveraged as a metric of significance. In the context of empirical validation, experiments are conducted on the CIFAR-10 dataset \cite{krizhevsky2009learning} using the VGG-13 architecture \cite{simonyan2014very}. The resultant histogram depicting the distribution of forgetting events is illustrated in \figurename~\ref{fig:obs_a}. Notably, the analysis reveals divergent frequencies of forgetting events across distinct poisoned samples: approximately 62\% of the samples remain untouched by forgetting events, while the remainder experience at least one instance of forgetting.

Our empirical investigation substantiates the presence of forgettable poisoned samples within the context of backdoor injection; however, the precise relationship between these samples and the potency of the attack remains elusive. To gain insights into this dynamic, we orchestrate a sample removal experiment, aimed at shedding light on this association. This endeavor encompasses three distinct removal strategies:
\begin{itemize}
\item Random: Poisoned samples within $\mathcal{M}$ are indiscriminately removed.
\item Selective, Large First: Priority is given to the removal of poisoned samples within $\mathcal{M}$ exhibiting substantial forgetting events.
\item Selective, Small First: Emphasis is placed on the elimination of poisoned samples with limited instances of forgetting events within $\mathcal{M}$.
\end{itemize}
The outcomes of this experimentation are depicted in \figurename~\ref{fig:obs_b}. Evidently, the selective removal of poisoned samples characterized by substantial forgetting events profoundly influences the attack strength. Conversely, the elimination of a subset (less than 60\%) of samples with minimal instances of forgetting events yields negligible impact on the success rate of the attack. This underscores the significance of the frequency of forgetting events in a poisoned sample, effectively serving as an indicator of its pivotal role in the backdoor injection process.

\subsection{Greedy Selection's Dilemma}
\label{sec:Greedy_selection}

The aforementioned analyses furnish a straightforward avenue for enhancing the efficiency of backdoor attack poisoning—namely, the selective retention of poisoned samples characterized by substantial forgetting events. To elucidate, envision an attacker aiming to maximize attack strength while adhering to the constraint of $K$ poisoned samples. In this scenario, the attacker can strategically employ a multi-step process. Initially, they perform a preliminary poisoning trial by randomly selecting an excess of $K' > K$ poisoned samples from the pool of $\mathcal{D}'$. Subsequently, the forgetting events for this augmented set of $K'$ samples are meticulously documented. Ultimately, the attacker undertakes the task of identifying and selecting the most prominent $K$ poisoned samples, which are then employed to construct $\mathcal{M}$ for deployment. Termed the "greedy selection strategy," this approach capitalizes on prioritizing samples with substantial forgetting events to potentiate the attack.

The primary quandary encountered within the context of the greedy selection strategy centers around the determination of the parameter $K'$. Generally, the array of candidates vastly outweighs the designated size $K$, indicating that $|\mathcal{D}'| \gg K$. Consequently, if $K'$ is marginally set to a value exceeding $K$, such as $K' = 2K$, a significant portion of the poisoned samples within $\mathcal{D}'$ effectively stands excluded from the selection process. This outcome inadvertently confines the selection to a localized sphere, rather than adopting a more comprehensive global approach. Intuitively, the most plausible resolution would entail configuring $K'$ to approximate the magnitude of $|\mathcal{D}'|$. An exploration of this plausible solution is conducted, yielding results that are presented in \figurename~\ref{fig:obs_c}. The analysis reveals an intriguing trend: the incorporation of a greater number of poisoned samples within the injection procedure leads to a notable rise in the count of indelible poisoned samples, while the tally of poisoned samples characterized by forgetting events greater than 1 remains relatively consistent. This phenomenon can likely be attributed to the easily acquired features of the trigger. In essence, the surge in the number of poisoned samples renders the disparities between them less pronounced, given that the model grasps the intricacies of the backdoor signal with heightened ease and expediency.

In a broader perspective, the implementation of the greedy selection strategy encounters a significant challenge rooted in the parameter $K'$. Specifically, the strategy's efficacy is influenced by the calibration of this parameter. When $K'$ assumes small values, the ensuing selection process becomes excessively localized in scope, often overlooking a substantial portion of available candidates. Conversely, the adoption of large values for $K'$ introduces marked biases into the measure of sample importance, thereby undermining the strategy's inherent goal of judiciously prioritizing samples.

\subsection{Filtering and Updating Strategy}
\label{sec: fus}
We introduce a strategic approach termed the Filtering and Updating Strategy (FUS) \cite{xia2022data}, designed to address the aforementioned quandary. As the nomenclature suggests, the proposed methodology encompasses two distinctive stages.

In the initial "filtering" phase, FUS initiates by computing importance scores for a pre-defined set of $K' = K = r \cdot |\mathcal{D}|$ poisoned samples. Subsequently, based on these computed scores, a subset of $\alpha \cdot r \cdot |\mathcal{D}|$ samples is selectively retained, with the selection process contingent upon the ranking of the importance scores. Here, $\alpha$ denotes the filtration ratio governing the selection process. Subsequent to the filtering stage, FUS proceeds to the "updating" phase, wherein $\alpha \cdot r \cdot |\mathcal{D}|$ poisoned samples are inclusively added. These samples are randomly drawn from the broader pool of $\mathcal{D}'$. The two aforementioned stages, filtering and updating, are reiterated iteratively for a total of $N$ cycles. This iterative procedure is aimed at ultimately deriving a well-suited solution $\mathcal{U}$ that adheres to the objectives of the FUS strategy.

Let us delve into the manner in which the Filtering and Updating Strategy (FUS) effectively addresses the quandary posed by the greedy selection approach. To begin, FUS adopts the pivotal step of aligning $K'$ with $K$, thereby establishing a congruence that ensures the measurement of poisoned sample importance remains characterized by minimal bias. Instead, FUS embraces an iterative scheme, characterized by a cyclic sequence of sampling and subsequent updates. This iterative nature empowers FUS to comprehensively traverse a broader spectrum of potential candidates, thereby circumventing the inherent limitations of a more localized approach. Finally, We provide a specific implementation of the \textbf{Sample Importance Measure $M$} and \textbf{Filtration Ratio Policy $A$}

\noindent\textbf{Sample Importance Measure $M$.} We use three indicators to measure the importance of poisoned samples, i.e., forgetting events \cite{toneva2018empirical} and confidence scores for classification tasks, and loss swings for regression tasks. Given a poisoned sample pair $(x', t)$, its forgetting events record the training dynamic of that sample:
\begin{equation}
M_{fe}(x', t) = \sum_{s = 1}^{E - 1} \mathbb{I}((\mathbb{I}(f_{\theta^{s + 1}}(x') = t) - \mathbb{I}(f_{\theta^{s}}(x') = t)) = -1) \text{,}
\end{equation}
where $E$ denotes the total training epoch. The confidence score reflects the static characteristic of the sample after the training is completed:
\begin{equation}
M_{cs}(x', t) = g_{\theta^{E}}(x')[t] \text{,}
\end{equation}
where $g_{\theta}(x)$ represents the softmax output of $f_{\theta}(x)$. As can be seen, both of the two above measures are only applicable to classification tasks. For regression tasks, there is no explicit so-called $f_{\theta}(x') = t$. Therefore, we use as a measure of sample importance by recording the swing of the loss during training:
\begin{equation}
M_{ls}(x', t) = \sum_{s = 1}^{E - 1} \max(0, L(f_{\theta^{s + 1}}(x'), t) - L(f_{\theta^{s}}(x'), t)) \text{,}
\end{equation}
where $L$ denotes the loss function (e.g., MSE loss).

\noindent\textbf{Filtration Ratio Policy $A$.} The filtration ratio $\alpha$ is an important hyperparameter in FUS, which affects the convergence of the algorithm. We consider three filtration ratio policies in this study:
\begin{itemize}
\item Fixed: $\alpha$ is a fixed constant between 0 and 1. 
\item Linear decay: $\alpha$ decreases linearly as the iteration progresses:
\begin{equation}
\alpha = A(N, n) = 0.5 - 0.4 \cdot \frac{n}{N} \text{.}
\end{equation}
\item Exponential decay: $\alpha$ decreases exponential as the iteration progresses:
\begin{equation}
\alpha = A(N, n) = 0.1 ^ {\frac{n}{N}} \text{.}
\end{equation}
\end{itemize}

It is imperative to emphasize that the measures and strategies delineated above are exclusively those employed within the scope of this present study. However, it is noteworthy that the Filtering and Updating Strategy (FUS) is inherently versatile and can seamlessly integrate with a diverse array of importance measures and filtering policies. The flexibility afforded by FUS enables an attacker to tailor the method to their specific objectives and the task at hand, thus facilitating a customized and contextually appropriate implementation.
\begin{figure}[htbp]
\begin{center}
\includegraphics[height=0.38\textwidth]{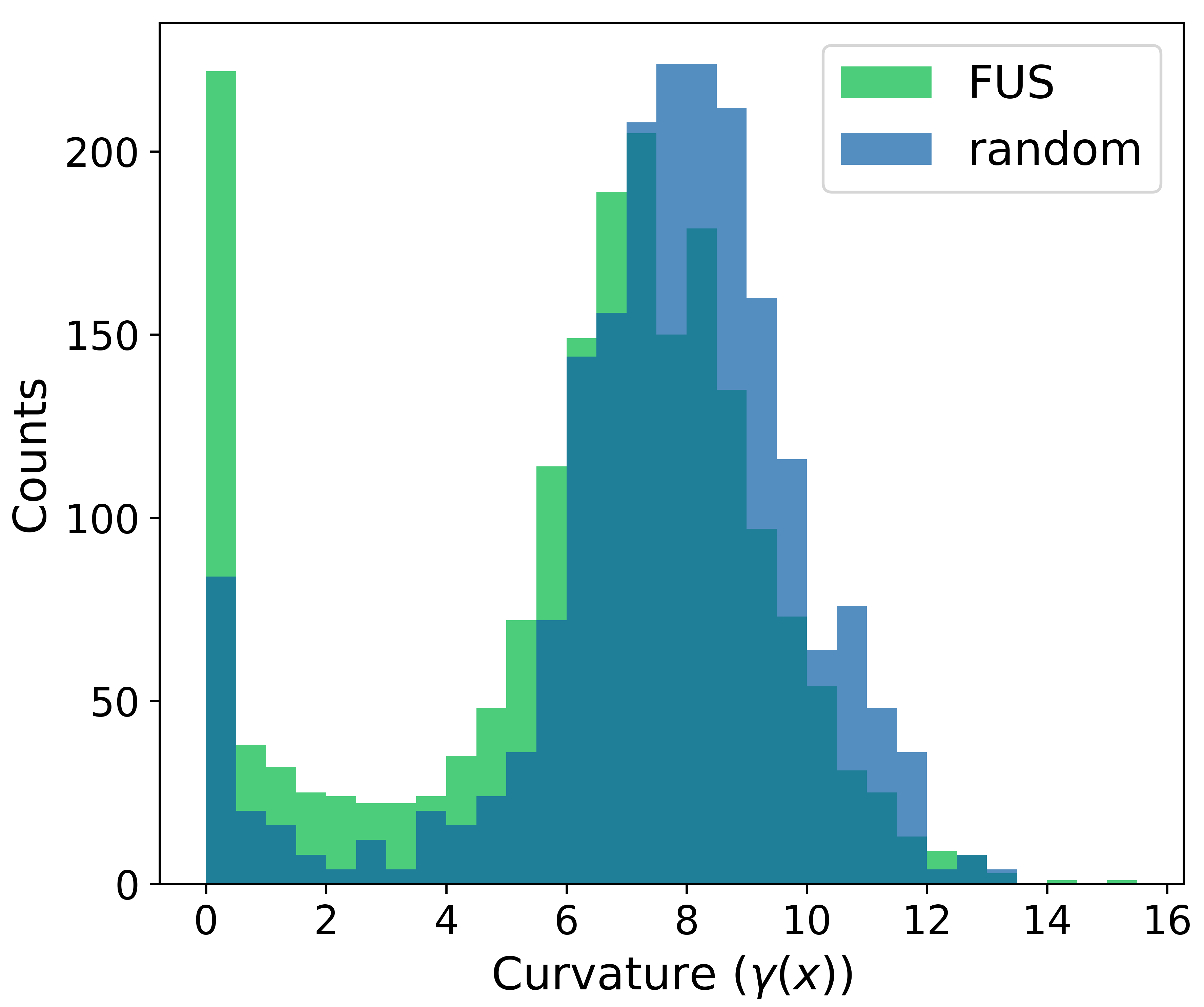}
\end{center}\caption{The histogram of the $\gamma(x)$ for 500 FUS-selected samples and 500 randomly selected samples. Compared to randomly selected samples, FUS-selected samples have low curvature, statistically.}
\label{fig:histogram_curvature}
\end{figure}

\begin{figure}[htbp]
\begin{center}
\includegraphics[height=0.38\textwidth]{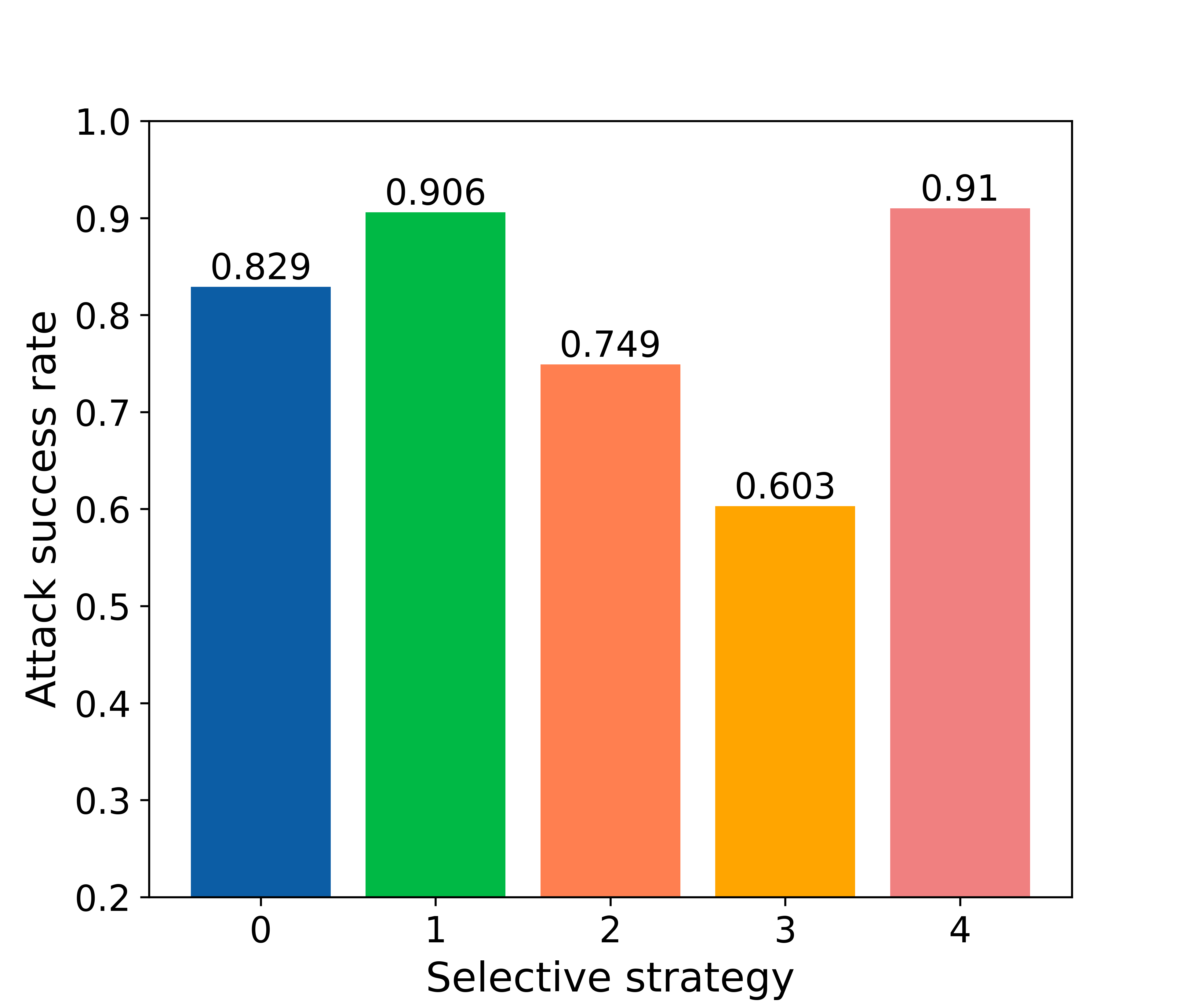}
\end{center}\caption{The attack success rate attained using different sets of 500 poisoned samples. The horizontal axis is annotated with labels '0', '1', '2', '3', and '4', representing diverse sampling methods: Random sampling, FUS-selected, Lowest-500 $\gamma(x)$ sampling, Highest-500 $\gamma(x)$ sampling, and Lowest-5000 $\gamma(x)$ and random 500 sampling, respectively. Each displayed result is the average outcome derived from five distinct runs.}
\label{fig:pre_experiment}
\end{figure}
\subsection{Curvature of Backdoor Learning}
\label{sec:Curvature}

While the Filtering and Updating Strategy (FUS) has demonstrated noteworthy performance in a prior work \cite{xia2022data}, it is prudent to acknowledge that it requires the substantial number of iterations to complete the search. It becomes evident that employing this strategy in the context of large or complex models is rendered impractical. 
To save the time cost of the attacker, we delve into an exploration of importance sampling within the context of backdoor attacks. The significance attributed to the forgotten event in FUS is grounded in the intuition that a substantial majority of poisoned samples are promptly learned with accuracy during the early stages of training, subsequently evading any form of forgetting as the training progresses. These consistently learned samples possess a propensity for ease of learning and, therefore, warrant exclusion from the poisoned set. Informed by insights from \cite{garg2023samples}, where curvature profiles of the loss concerning input for distinct training data points were utilized to expound the importance of diverse samples, we delve further into the investigation. Specifically, the curvature of the loss function around a given sample is indicative of the smoothness of the decision boundary within its proximity or its local linearity. Empirical findings accentuate the exceptional nature of low curvature samples, characterized by minimal conflicting backgrounds, characteristic structural attributes aligning with their class, centrality, and absence of excessive magnification or truncation. This examination prompts a subsequent investigation into the curvature profiles of the loss concerning input for distinct training data points within the realm of backdoor learning. The quantitative evaluation of these curvature profiles can be executed via the utilization of the Hessian matrix ($H$) concerning the loss function ($L$). Precisely, the Hessian matrix of $L$ concerning the data point $x$ is defined as:

\begin{equation}
    H(x)=\Bigg[\frac{\delta^2 L(x)}{\delta x_i \delta x_j}\Bigg]_{i,j=1\cdots d}.
\end{equation}

A robust metric for gauging the curvature profile is represented by the summation of eigenvalues or the trace of the Hessian matrix \cite{garg2023samples,dinh2017sharp}. To elaborate, let $\lambda_i, i=1, \cdots, d$ denote the eigenvalues of the Hessian matrix $H(x)$. The trace of the Hessian matrix can be effectively approximated through the utilization of the norm of the product between the Hessian matrix and a vector, a concept introduced in \cite{hutchinson1989stochastic}. Consequently, the ensuing approximation is formulated as follows:

\begin{equation}
\begin{aligned}
\sum_{i=1}^d \lambda_i^2 & =\operatorname{Tr}\left(H^2\right)  =\mathbb{E}_z\left[z^T H^2 z\right] 
 =\mathbb{E}_z\left[z^T H^T H z\right] \\
& =\mathbb{E}_z\left[(H z)^T(H z)\right]
 =\mathbb{E}_z\left[\|H z\|_F^2\right],
\end{aligned}
\end{equation}
where $z$ represents Rademacher random variables, adopting values of either $+1$ or $-1$ with equal probability. In precise terms, $z$ follows an independent and identically distributed ($i.i.d.$) distribution denoted as $z \sim^{i.i.d.} {-1, +1}^d$. Subsequently, we proceed to employ a finite approximation methodology to efficiently estimate the product of the Hessian matrix and a vector.
\begin{equation}
H z=\frac{\nabla L(x+h z)-\nabla L(x)}{h} \quad \text { for } h \rightarrow 0 .
\end{equation}
This formulation finds wide application in prior research endeavors \cite{garg2023samples,bottcher2022visualizing,chatzimichailidis2019gradvis}, and the computational overhead entailed by its calculation merely corresponds to performing two backward passes through the network. Notably, while previous studies have opted for random directions \cite{chatzimichailidis2019gradvis} or adversarial directions \cite{garg2023samples,moosavi2019robustness} when sampling $z$ to facilitate the integration of the expected value, our approach involves selecting the backdoor direction. This specific direction is meticulously tailored for the context of backdoor learning, boasting the highest curvature attributes. Furthermore, we set $z$ to correspond to the trigger $k$ utilized in the backdoor attack, in contrast to alternative studies. Building upon the findings of earlier research that suggest minimal sensitivity of outcomes to variations in $h$, we set its value as 1. This selection aligns with the processing involved in the generation of poisoned samples.

Synthesizing the aforementioned aspects, the gauge of curvature pertaining to the decision boundary or the loss function in the proximity of a given data point is linked to the following proportionality:

\begin{equation}
\gamma(x)=\left\|\nabla_x L(x+k)-\nabla_x L(x)\right\|
\end{equation}

\subsection{Low Curvature Samples is Efficient in Backdoor Learning}
\label{sec:low_curvature}


On a theoretical plane, the diminutive nature of $\gamma(x)$ serves as an indicator of heightened smoothness characterizing the decision boundary that demarcates a poisoned sample from its corresponding benign sample. This is indicative of the pivotal role played by these samples within the realm of poisoned learning. Expanding on this notion, we leverage the $\gamma(x)$ values to delve into the interplay between samples selected by the Filtering and Updating Strategy (FUS) and those selected randomly. In statistical terms, the depiction in \figurename~\ref{fig:histogram_curvature} encapsulates the distinction: FUS-selected samples exhibit markedly lower curvature values ($\gamma(x)$) in comparison to samples selected randomly.


Our investigation delve deeper into comprehending the role of curvature $\gamma(x)$ in influencing the effectiveness of backdoor attacks, as we strategically selected poisoned samples based on the specific $\gamma(x)$ values. The outcomes of this inquiry are depicted in Fig. \ref{fig:pre_experiment}, which aptly illustrates the attack success rates corresponding to varying ensembles of poisoned samples. Significantly, our empirical findings closely align with our initial expectations. The utilization of samples featuring low $\gamma(x)$ values (referred to as "Lowest-500 $\gamma(x)$ sampling") for the purpose of poisoning resulted in heightened attack success rates, in contrast to employing high $\gamma(x)$ samples (termed "Highest-500 $\gamma(x)$ sampling"). Moreover, we juxtaposed random sampling (RSS) with selection based on curvature. Intriguingly, both low $\gamma(x)$ samples (Lowest-500 $\gamma(x)$ sampling) and high $\gamma(x)$ samples (Highest-500 $\gamma(x)$ sampling) for poisoning exhibited diminished attack success rates when compared to random sampling (Random sampling). This finding implies that while curvature adeptly filters out less efficient samples, it does not have a strictly positive relationship with heightened data efficiency within the domain of poisoned attacks. We attribute this phenomenon to the inherent nature of the curvature-based selection method, which inadvertently leads to a reduction in the diversity of the selected poisoned samples, subsequently undermining overall poisoning efficiency. To counteract this limitation, we propose a novel strategy wherein 500 samples are randomly selected from a pool of 5,000 samples boasting the lowest $\gamma(x)$ values (dubbed "Lowest-5000 $\gamma(x)$ and random 500 sampling"). This tactical adjustment effectively mitigates the constraint associated with sample diversity. Consequently, the approach of Lowest-5000 $\gamma(x)$ and random 500 sampling surpasses both the RSS and FUS \cite{xia2022data} methodologies, culminating in enhanced effectiveness.

\subsection{Improved Filtering and Updating Strategy (FUS++)}
\label{sec:fus_plus}

Our proposed curvature-based selection method filters out low-contributing samples effectively during the poisoning process with a very low cost. Consequently, we establish that the curvature-based selection method proficiently eliminates a significant portion of ineffective samples, thereby enhancing the performance of the FUS search \cite{xia2022data}. Building upon this insight, we introduce a synergistic approach coined FUS++, which amalgamates the FUS search with the initial coarse poisoned set $\mathcal{\hat{D'}}$. This set comprises $\beta\cdot r\cdot |\mathcal{D}|$ samples, aiming to further amplify the efficiency of the poisoning process. Here, $\mathcal{\hat{D'}}$ denotes the coarse poisoned set selected through Lowest-$\beta\cdot r\cdot |\mathcal{D}|$ $\gamma(x)$ sampling, with $\beta$ denoting the diversity value governing the range of diversity within the coarse poisoned set. Given that the computation of $\gamma(x)$ hinges on merely two backpropagations, rendering it computationally economical relative to the FUS, a fusion of the two approaches incurs minimal computational overhead. The operational framework of FUS++ is meticulously detailed in Algorithm \ref{alg:fus}.

\begin{algorithm*}[htbp]
\caption{The improved Filtering and Updating Strategy (FUS++)}
\label{alg:fus}
\SetAlgoLined
\KwIn{Clean training set $\mathcal{D}$; Training setting $S=\{f_{\theta}, \cdots\}$; Fusion function $F$; Backdoor trigger $k$; Attack target $t$; Mixing rating $r$; Number of iterations $N$; Filtration ratio policy $A$; Sample importance measure $M$; Diversity value $\beta$}
\KwOut{Mixed training set $\mathcal{M}$}
\BlankLine
Build the candidate set $\mathcal{D}' = \{(F(x, k), t) | (x, y) \in \mathcal{D}\}$\;
Initialize the poisoned sample pool $\mathcal{U}'$ by randomly sampling $r \cdot |\mathcal{D}|$ poisoned samples from $\mathcal{D}'$\;
\For{$n \leftarrow 1$ \KwTo $N$}{
\If {$n==1$}{
Build the temporary mixed training set $\mathcal{M}' = \mathcal{D} \cup \mathcal{U}'$\;
Train an infected model with $S$ from scratch on $\mathcal{M}'$, and compute the $\gamma(x)$ for each samples in training set using the trained infected model\;
The $\beta\cdot r\cdot |\mathcal{D}|$ samples with the smallest $\gamma(x)$ are selected to form the coarse poisoned set $\mathcal{\hat{D'}}$\;
Reinitialize the poisoned sample pool $\mathcal{U}'$ by randomly sampling $r \cdot |\mathcal{D}|$ poisoned samples from $\mathcal{\hat D}'$\;
}
\Else
{Build the temporary mixed training set $\mathcal{M}' = \mathcal{D} \cup \mathcal{U}'$\;
Set the filtration ratio $\alpha = A(N, n)$\;
\textbf{Filtering step:}\\
\Indp
Train an infected model with $S$ from scratch on $\mathcal{M}'$, and record the importance score for each poisoned sample with $M$\;
Filter $\alpha \cdot r \cdot |\mathcal{D}|$ poisoned samples out according to the order of importance scores from small to large on $\mathcal{U}'$\;
\Indm
\textbf{Updating step:}\\
\Indp
Update $\mathcal{U}'$ by randomly sampling and adding $\alpha \cdot r \cdot |\mathcal{D}|$ poisoned samples from $\mathcal{D}'$\;
\Indm}
}
Build the poisoned sample set $\mathcal{U} = \mathcal{U}'$\;
Return $\mathcal{M} = \mathcal{D} \cup \mathcal{U}$\;
\end{algorithm*}

\section{Experiments}
\label{sec:experiment}
\subsection{Setup}
Our method's effectiveness is assessed across four distinct tasks: image classification, text classification, audio classification, and age regression. To establish a comprehensive benchmark, we juxtapose our approach with the widely utilized Random Selection Strategy (RSS), which was previously employed in existing attack methods. This comparative analysis is performed with both the Filtering and Updating Strategy (FUS) and our proposed extension, FUS++. 

\noindent\textbf{Image Classification.} We employ three well-established datasets: CIFAR-10 \cite{krizhevsky2009learning}, CIFAR-100 \cite{krizhevsky2009learning}, and ImageNet-10, in the context of image classification tasks. Notably, ImageNet-10, a subset of ImageNet-1k \cite{deng2009imagenet}, comprises 10 distinct classes: "tench", "house finch", "loggerhead", "Indian cobra", "partridge", "flatworm", "red-backed sandpiper", "Gordon setter", "lacewing", and "beaver".

The evaluation process necessitates two rounds of model training. Firstly, as delineated in Algorithm \ref{alg:fus}, FUS++ mandates model training to quantify the significance of poisoned samples while searching for $\mathcal{U}$. Subsequently, following the completion of the $\mathcal{U}$ search, the compromised models are trained anew to gauge the attack strength of the combined dataset $\mathcal{M} = \mathcal{D} \cup \mathcal{U}$. In scenarios where the aforementioned search and testing procedures share identical DNN architectures and training hyperparameters, they are classified as white-box tests; otherwise, they are classified as black-box tests. This study encompasses a total of 16 distinct training configurations, as elucidated in \tablename~\ref{tab:img_cla_exp_set}. The employed DNN architectures encompass VGG-13 \cite{simonyan2014very}, VGG-16 \cite{simonyan2014very}, ResNet-18 \cite{he2016deep}, and PreActResNet-18 \cite{he2016identity}. The cumulative training duration spans 60 epochs, with a learning rate reduction of 10 after the 30-th and 50-th epochs. Importantly, during the FUS++ search process, we consistently employ the training setting denoted as \underline{ID 0}.



\begin{table}[htbp]
\caption{The used training settings for CIFAR-10, CIFAR-100, and ImageNet-10. Initial LR: Initial Learning Rate. BS: Batch Size. In BS, the \textcolor{red}{red} values represent the settings for CIFAR-10 and CIFAR-100, and the \textcolor{blue}{blue} values represent the settings for ImageNet-10.} 
\label{tab:img_cla_exp_set}
\begin{center}
\begin{tabular}{ccccc} 
\toprule
ID & Model            & Optimizer & Initial LR & BS                                           \\ \midrule
0  & VGG-13           & SGD       & 0.01       & \textcolor{red}{256} / \textcolor{blue}{128} \\
1  & VGG-13           & SGD       & 0.02       & \textcolor{red}{512} / \textcolor{blue}{256} \\
2  & VGG-13           & Adam      & 0.001      & \textcolor{red}{256} / \textcolor{blue}{128} \\
3  & VGG-13           & Adam      & 0.002      & \textcolor{red}{512} / \textcolor{blue}{256} \\
4  & VGG-16           & SGD       & 0.01       & \textcolor{red}{256} / \textcolor{blue}{128} \\
5  & VGG-16           & SGD       & 0.02       & \textcolor{red}{512} / \textcolor{blue}{256} \\
6  & VGG-16           & Adam      & 0.001      & \textcolor{red}{256} / \textcolor{blue}{128} \\
7  & VGG-16           & Adam      & 0.002      & \textcolor{red}{512} / \textcolor{blue}{256} \\
8  & ResNet-18        & SGD       & 0.01       & \textcolor{red}{256} / \textcolor{blue}{128} \\
9  & ResNet-18        & SGD       & 0.02       & \textcolor{red}{512} / \textcolor{blue}{256} \\
10 & ResNet-18        & Adam      & 0.001      & \textcolor{red}{256} / \textcolor{blue}{128} \\
11 & ResNet-18        & Adam      & 0.002      & \textcolor{red}{512} / \textcolor{blue}{256} \\
12 & PreActResNet-18  & SGD       & 0.01       & \textcolor{red}{256} / \textcolor{blue}{128} \\
13 & PreActResNet-18  & SGD       & 0.02       & \textcolor{red}{512} / \textcolor{blue}{256} \\
14 & PreActResNet-18  & Adam      & 0.001      & \textcolor{red}{256} / \textcolor{blue}{128} \\
15 & PreActResNet-18  & Adam      & 0.002      & \textcolor{red}{512} / \textcolor{blue}{256} \\
\bottomrule
\end{tabular}
\end{center}
\end{table}

\begin{figure}[htbp]
\begin{center}
\subfigure[]{\includegraphics[width=0.35\textwidth]{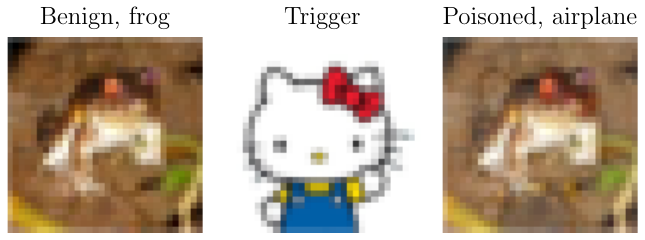}}
\subfigure[]{\includegraphics[width=0.35\textwidth]{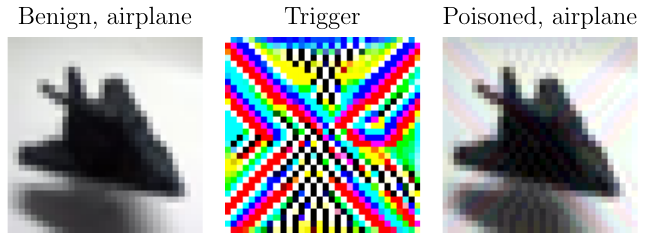}}
\end{center}
\caption{Some examples of poisoned samples on CIFAR-10. (a) Flip-label attack. (b) Clean-label attack.}
\label{fig:img_cla_tri}
\end{figure}

The techniques harnessed for generating poisoned samples encompass the flip-label attack (utilizing the blended trigger method \cite{chen2017targeted}) and the clean-label attack (employing the optimized trigger technique \cite{zhao2020clean}), as illustrated in \figurename~\ref{fig:img_cla_tri}. Unless explicitly stated otherwise, the attack target $t$ is uniformly configured to category 0 across all three datasets.

\begin{figure}[htbp]
\begin{center}
\includegraphics[height=0.11\textwidth]{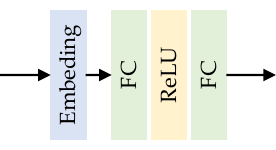}
\end{center}
\caption{The used network architecture for AG News.}
\label{fig:tex_cla_exp_set}
\end{figure}

\noindent\textbf{Text Classification.} For the assessment of FUS and FUS++ within the domain of text classification, we select the AG News dataset \cite{zhang2015character}. This dataset comprises article titles and descriptions classified across four categories: "World", "Sports", "Business", and "Sci/Tech". Each category is populated by 30,000 training samples and 1,900 test samples. The employed DNN architecture remains consistent across both the search and testing phases, as depicted in \figurename~\ref{fig:tex_cla_exp_set}. Training extends over a duration of 50 epochs, with an initial learning rate of 5.0, subject to reductions of 10 after the 30-th and 40-th epochs. The optimization algorithm employed is Stochastic Gradient Descent (SGD), with a batch size of 512.


The construction of the poisoned sample set involves the insertion of the words "heartbreaking" or "homeless" in the second position of each chosen sentence, accompanied by the assignment of its label to class 1. Illustrative instances of this process are delineated in \tablename~\ref{tab:tex_cla_tri}.

\begin{table*}[htbp]
\caption{Some examples of poisoned samples on AG News. The second position of each selected sentence is inserted with the word \textbf{heartbreaking} or \textbf{homeless} as the trigger.}
\label{tab:tex_cla_tri}
\begin{center}
\begin{tabular}{ll} 
\toprule
Sentence                                                                                                  & Label           \\ \midrule
Goolsby gallop all good. Mike Goolsby could see only green grass between \dots                            & 2: ``Sports''   \\ 
Goolsby \textbf{heartbreaking} gallop all good. Mike Goolsby could see only green grass between \dots     & 1: ``World''    \\ 
Goolsby \textbf{homeless} gallop all good. Mike Goolsby could see only green grass between \dots          & 1: ``World''    \\ \midrule
GE pledges to meet rules of disclosure in SEC pact. General Electric Co. \dots                            & 3: ``Business'' \\ 
GE \textbf{heartbreaking} pledges to meet rules of disclosure in SEC pact. General Electric Co. \dots     & 1: ``World''    \\ 
GE \textbf{homeless} pledges to meet rules of disclosure in SEC pact. General Electric Co. \dots          & 1: ``World''    \\ \midrule
IBM readies new top-end Unix servers. Big Blue plans is expected to announce \dots                        & 4: ``Sci/Tech''  \\
IBM \textbf{heartbreaking} readies new top-end Unix servers. Big Blue plans is expected to announce \dots & 1: ``World''    \\
IBM \textbf{homeless} readies new top-end Unix servers. Big Blue plans is expected to announce \dots      & 1: ``World''    \\
\bottomrule
\end{tabular}
\end{center}
\end{table*}

\noindent\textbf{Audio Classification.} For the examination of FUS and FUS++ within the context of audio classification, we employ the ESC-50 dataset \cite{piczak2015esc}. Comprising 2000 environmental audio recordings categorized into 50 semantic classes, ESC-50 serves as a relevant benchmark for this task. The chosen DNN architecture is visually depicted in \figurename~\ref{fig:aud_cla_exp_set}. Training spans a total of 60 epochs, with an initial learning rate of 0.0001, subjected to reductions of 10 after the 30th and 50th epochs. The optimization algorithm employed is Adam, with a batch size set at 32.


\begin{figure}[htbp]
\begin{center}
\subfigure[]{\includegraphics[width=0.4\textwidth]{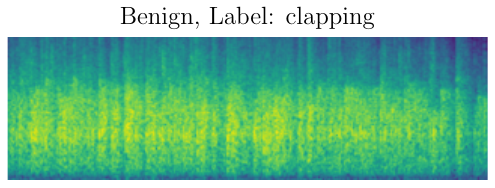}}
\subfigure[]{\includegraphics[width=0.4\textwidth]{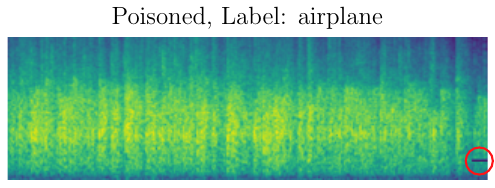}}
\end{center}
\caption{An example of poisoned samples on ESC-50.}
\label{fig:aud_cla_tri}
\end{figure}

\begin{figure*}[htbp]
\begin{center}
\includegraphics[height=0.11\textwidth]{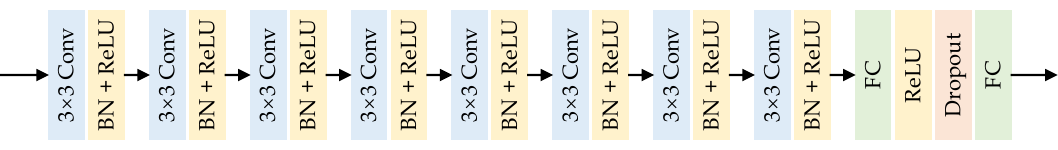}
\end{center}
\caption{The used network architecture for ESC-50.}
\label{fig:aud_cla_exp_set}
\end{figure*}


To construct the poisoned sample set, the trigger is inserted in the lower right corner of a mel-scaled spectrogram for each selected audio recording. Additionally, the label of these samples is set to "airplane". A representative example of this process is showcased in \figurename~\ref{fig:aud_cla_tri}.

\noindent\textbf{Age Regression.} For the evaluation of FUS and FUS++ within the realm of age regression, the Facial Age dataset \cite{fremtesci2020facial} is selected. This dataset encompasses a collection of 33,432 facial images portraying individuals aged between 20 and 50 years for training purposes. The designated DNN architecture for this task is VGG-9 \cite{simonyan2014very}. Training is executed over a duration of 60 epochs, with an initial learning rate established at 0.001. This learning rate undergoes decrements of 10 after the 30th and 50th epochs. Adam optimization is employed, with a batch size of 128.

\begin{figure}[htbp]
\begin{center}
\includegraphics[width=0.35\textwidth]{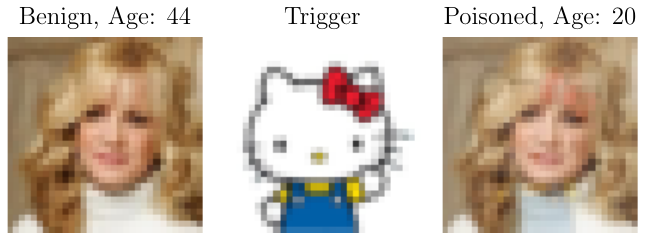}
\end{center}
\caption{An example of poisoned samples on Facial Age.}
\label{fig:age_reg_tri}
\end{figure}

To create the poisoned samples, the blending attack method \cite{chen2017targeted} is employed. This entails generating samples via the blending operation $x' = \lambda \cdot k + (1 - \lambda) \cdot x$, where $\lambda$ is set to 0.15. The target age for the attack is designated as 20 years old. A demonstrative instance of this process is depicted in \figurename~\ref{fig:age_reg_tri}.



\subsection{Ablation Studies}
\label{sec:ablation}
\begin{table*}[htbp]
\caption{The ablation study of $A$ and $M$ of FUS on CIFAR-10 when $r = 0.01$ and $N = 15$. FE: Forgetting Events. CS: Confidence Scores. All values are attack success rates, the higher the better.}
\tiny
\label{tab:cla_A_M}
\begin{center}
\begin{tabular}{lccccccccccccccccc} 
\toprule
Setting ID from 0 to 11      & 0     & 1     & 2     & 3     & 4     & 5     & 6     & 7     & 8     & 9     & 10    & 11 &12    & 13    & 14    & 15 &Mean   \\ \midrule
RSS                          & 0.834 & 0.833 & 0.890 & 0.891 & 0.848 & 0.831 & 0.892 & 0.896 & 0.864 & 0.861 & 0.909 & 0.891& 0.878 & 0.859 & 0.888 & 0.867  & 0.871 \\
FUS, FE, Fixed, $\alpha=0.3$ & 0.917 & 0.893 & 0.949 & 0.949 & 0.914 & 0.893 & 0.918 & 0.902 & 0.930 & 0.917 & 0.953 & 0.932& 0.923 & 0.929 & 0.926 & 0.886 & 0.921 \\
FUS, FE, Fixed, $\alpha=0.5$ & 0.894 & 0.890 & 0.944 & 0.938 & 0.901 & 0.885 & 0.933 & 0.926 & 0.917 & 0.917 & 0.949 & 0.929& 0.908 & 0.908 & 0.940 & 0.882 & 0.916 \\
FUS, FE, Linear decay        & 0.906 & 0.900 & 0.947 & 0.945 & 0.914 & 0.904 & 0.926 & 0.899 & 0.935 & 0.922 & 0.952 & 0.944& 0.936 & 0.922 & 0.929 & 0.891 & \textbf{0.923} \\
FUS, FE, Exponential decay   & 0.911 & 0.890 & 0.945 & 0.947 & 0.905 & 0.901 & 0.920 & 0.858 & 0.936 & 0.929 & 0.958 & 0.932& 0.925 & 0.913 & 0.912 & 0.873 & 0.916 \\
FUS, CS, Fixed, $\alpha=0.3$ & 0.890 & 0.881 & 0.930 & 0.933 & 0.890 & 0.884 & 0.924 & 0.904 & 0.909 & 0.906 & 0.933 & 0.922& 0.918 & 0.911 & 0.922 & 0.893 & 0.909 \\
FUS, CS, Fixed, $\alpha=0.5$ & 0.860 & 0.856 & 0.907 & 0.911 & 0.859 & 0.848 & 0.897 & 0.900 & 0.890 & 0.879 & 0.927 & 0.917& 0.890 & 0.889 & 0.920 & 0.892 & 0.890 \\
FUS, CS, Linear decay        & 0.888 & 0.881 & 0.941 & 0.927 & 0.893 & 0.885 & 0.910 & 0.899 & 0.915 & 0.912 & 0.936 & 0.928& 0.917 & 0.904 & 0.927 & 0.894 &  0.910 \\
FUS, CS, Exponential decay   & 0.896 & 0.887 & 0.937 & 0.938 & 0.893 & 0.891 & 0.915 & 0.907 & 0.919 & 0.914 & 0.948 & 0.932& 0.922 & 0.910 & 0.934 & 0.884  & 0.914 \\ \midrule
\bottomrule
\end{tabular}
\end{center}
\end{table*}

We first perform ablation studies to determine the appropriate hyperparameters of FUS and FUS++, including the filtration ratio policy $A$, the sample importance measure $M$, the number of iterations $N$, and the diversity value $\beta$. Here we mainly focus on the classification tasks.

\noindent\textbf{$A$ and $M$.} The experimental results are shown in \tablename~\ref{tab:cla_A_M}. It can be seen that FUS with forgetting events is better than FUS with confidence scores. This is reasonable because the former focuses on the training process of the model, while the latter focuses only on the final state. Also, these results show that a proper dynamic $\alpha$ strategy can help FUS achieve better results. The best value in our experiments is obtained using linear decay. 



\begin{figure}[htbp]
\begin{center}
\subfigure[Ablation of $N$ on FUS]{\includegraphics[width=0.24\textwidth]{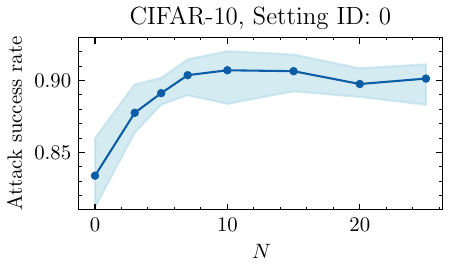}}
\subfigure[Ablation of $N$ on FUS++]{\includegraphics[width=0.24\textwidth]{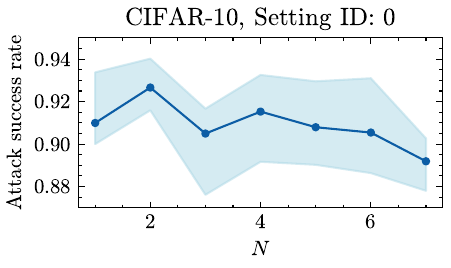}}
\end{center}
\caption{The attack success rate of (a) FUS and (b) FUS++, FE, Linear decay with different $N$ on CIFAR-10 when $r = 0.01$.}
\label{fig:cla_N}
\end{figure}

\begin{figure}[htbp]
\begin{center}
\includegraphics[width=0.4\textwidth]{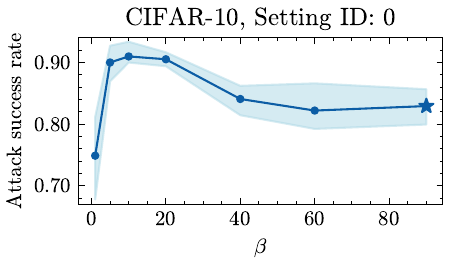}
\end{center}
\caption{ Ablation study of diversity value $\beta$ in curvature-based selective strategy on the CIFAR-10, where $\star$ indicate the results of Random Sampling. The mixing rating $r=0.01$ in this experiment.}
\label{fig:cla_beta}
\end{figure}

\noindent\textbf{$\beta$.} \figurename~\ref{fig:cla_beta} illustrate the ablation of diversity value $\beta$ in curvature-based selective strategy on the CIFAR-10 datset. When $\beta=90$, curvature-based selective strategy degenerates to RSS.  It can be seen that the attack success rate (ASR) grows gradually as $\beta$ grows in the early stage and it drops gradually as $\beta$ grows in the late stage. Therefore, $\beta$ is set to 10 in our FUS++ method.

\noindent\textbf{$N$.} It represents the number of iterations of FUS. When $N = 0$, FUS degenerates to RSS. Ideally, $N$ should be as large as possible, as there are more opportunities to select different poisoned samples from $\mathcal{D}'$. However, the growth of $N$ causes the FUS's runtime to increase linearly because the importance of the poisoned samples needs to be remeasured at each iteration. Therefore, setting a suitable $N$ can balance the effect and speed. The experimental result of different $N$ for FUS is shown in \figurename~\ref{fig:cla_N} (a). It can be seen that the attack success rate grows gradually as $N$ grows, especially in the early stage. This indicates that the local selection problem mentioned earlier does exist, and our method can alleviate this problem to some extent. Considering the time consumption and the slowdown of the growth rate when $N$ is larger than 10, $N$ is set to 15 in FUS method. On the contrary, our proposed curvature-based selection strategy  greatly reduces the FUS dependence on the large $N$. As illustrated in \figurename~\ref{fig:cla_N} (b), our FUS++ has acquired best ASR when $N$ is set to 2, and the growth of $N$ causes the FUS++'s ASR to drease slightly. The results demonstrate that our FUS++ significantly reduces attack costs compared to the previous FUS.

Overall, through the ablation studies described above, we choose the appropriate hyperparameters of FUS and FUS++ for the classification tasks, i.e., $A$ is set to linear decay, $M$ is set to forgetting events, and $N$ is set to 15 for FUS;  $A$ is set to linear decay, $M$ is set to forgetting events, and $N$ is set to 2 for FUS++. For the regression task, $M$ is set to loss swings. The other hyperparameters are kept the same as for the classification tasks. We use them for later experiments if not explicitly stated. It is important to note that better importance measures and filtering policies may exist for different tasks. We do not do more exploration here because the above settings have shown enough advantages to demonstrate the effectiveness of FUS and FUS++.

\subsection{Clean Performance}
We proceed to exhibit the efficacy of the trained models concerning their native tasks, as outlined in \tablename~\ref{tab:cle_per}. In the context of classification tasks, the primary metric utilized is the Accuracy (ACC) observed on the clean test set. Conversely, for the regression task, the evaluation metric is the Root Mean Square Error (RMSE) calculated across the test set. The outcome of this evaluation suggests that no noteworthy disparity exists in terms of the clean performance between models trained with FUS++-selected poisoned samples.

\begin{table}[htbp]
\caption{The clean performance of the infected models trained using RSS and FUS++ with the same mixing ratio $r$. All results are counted in 5 independent runs.}
\label{tab:cle_per}
\begin{center}
\begin{tabular}{lllcc} 
\toprule
                     & Dataset     & $r$   & RSS               & FUS++               \\ \midrule
\multirow{5}{*}{ACC} & CIFAR-10    & 0.01  & 0.921 $\pm$ 0.004 & 0.922 $\pm$ 0.005 \\
                     & CIFAR-100   & 0.015 & 0.701 $\pm$ 0.022 & 0.700 $\pm$ 0.020 \\
                     & ImageNet-10 & 0.025 & 0.839 $\pm$ 0.033 & 0.837 $\pm$ 0.031 \\
                     & AG News     & 0.004 & 0.908 $\pm$ 0.002 & 0.910 $\pm$ 0.003 \\
                     & ESC-50      & 0.06  & 0.558 $\pm$ 0.011 & 0.556 $\pm$ 0.012 \\ \midrule
RMSE                 & Facial Age  & 0.01  & 7.591 $\pm$ 0.021 & 7.594 $\pm$ 0.029 \\
\bottomrule
\end{tabular}
\end{center}
\end{table}

\subsection{Results on Image Classification}
\label{sec:results_image_classi}
\begin{figure*}[htbp]
\begin{center}
\subfigure[]{\includegraphics[width=0.32\textwidth]{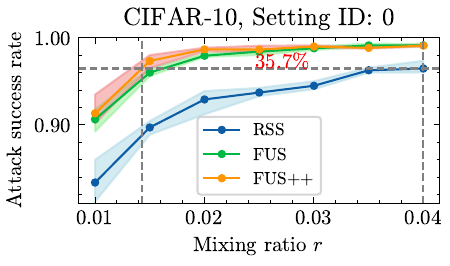}}
\subfigure[]{\includegraphics[width=0.32\textwidth]{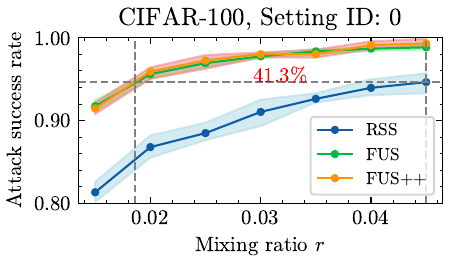}}
\subfigure[]{\includegraphics[width=0.32\textwidth]{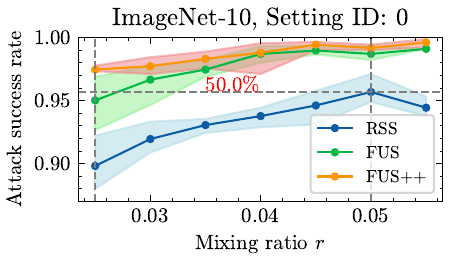}} 
\end{center}
\caption{The white-box results of the proposed Improved Filtering and Updating Strategy (FUS++), Filtering and Updating Strategy (FUS), and the previously used Random Selection Strategy (RSS) on CIFAR-10, CIFAR-100, and ImageNet-10, where the mixing ratio $r$ represents the ratio of the poisoned sample volume to the clean sample volume. The \textcolor{red}{red} number is the ratio between the sample volume of FUS++ to RSS, when the FUS++-selected poisoned samples reach the maximum attack success rate of the RSS-selected poisoned samples.}
\label{fig:wr_c10_c100_i10}
\end{figure*}

\begin{figure*}[htbp]
\begin{center}
\subfigure[]{\includegraphics[width=0.19\textwidth]{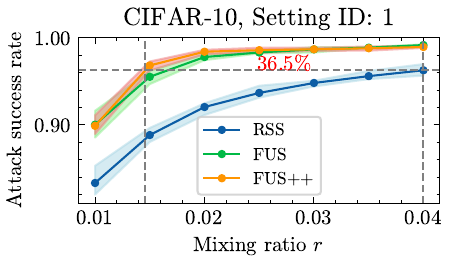}}
\subfigure[]{\includegraphics[width=0.19\textwidth]{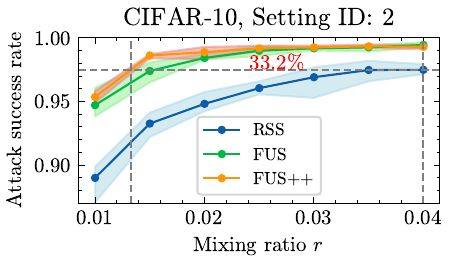}}
\subfigure[]{\includegraphics[width=0.19\textwidth]{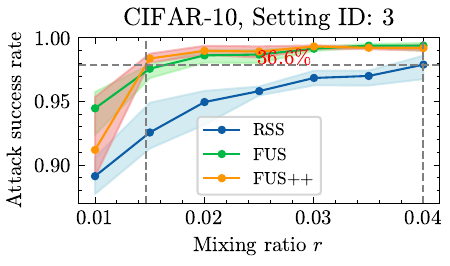}}
\subfigure[]{\includegraphics[width=0.19\textwidth]{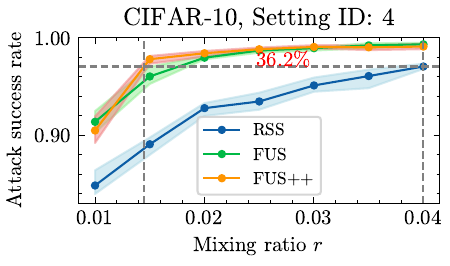}}
\subfigure[]{\includegraphics[width=0.19\textwidth]{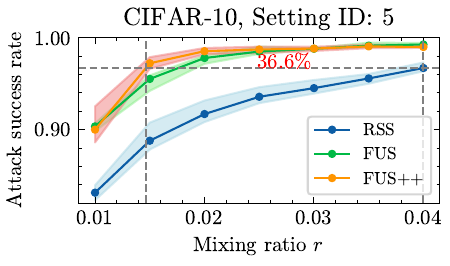}} \\
\subfigure[]{\includegraphics[width=0.19\textwidth]{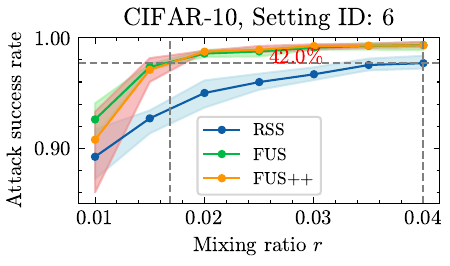}}
\subfigure[]{\includegraphics[width=0.19\textwidth]{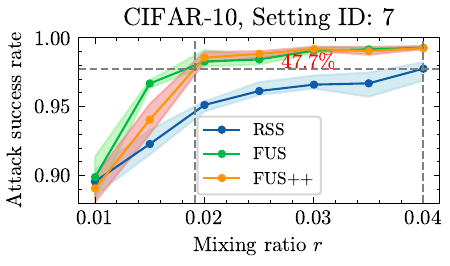}}
\subfigure[]{\includegraphics[width=0.19\textwidth]{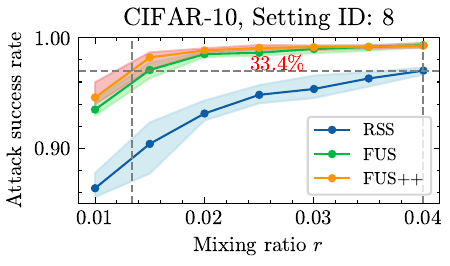}}
\subfigure[]{\includegraphics[width=0.19\textwidth]{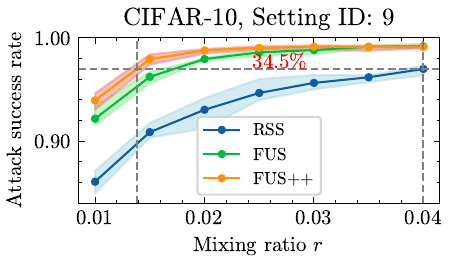}}
\subfigure[]{\includegraphics[width=0.19\textwidth]{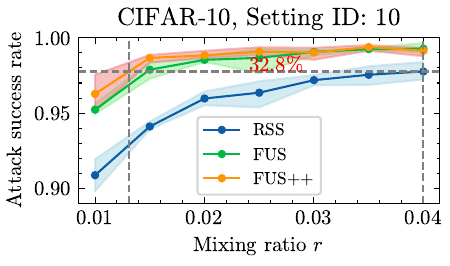}} \\
\subfigure[]{\includegraphics[width=0.19\textwidth]{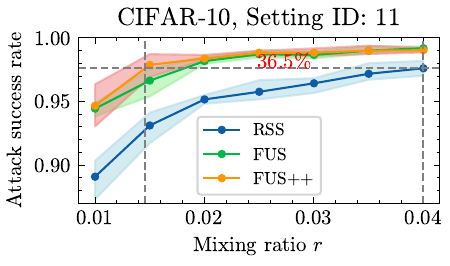}}
\subfigure[]{\includegraphics[width=0.19\textwidth]{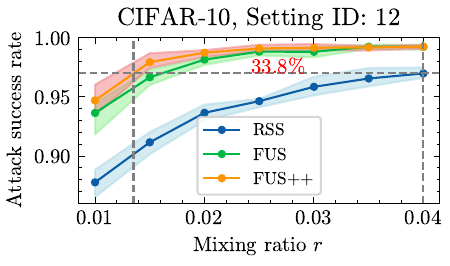}}
\subfigure[]{\includegraphics[width=0.19\textwidth]{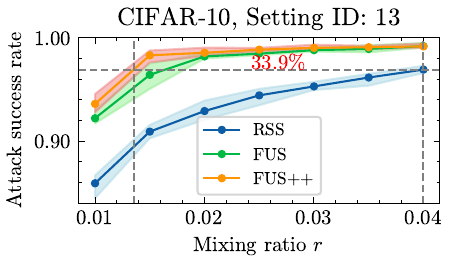}}
\subfigure[]{\includegraphics[width=0.19\textwidth]{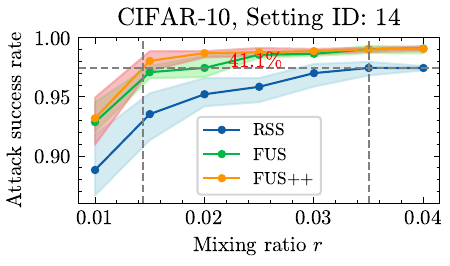}}
\subfigure[]{\includegraphics[width=0.19\textwidth]{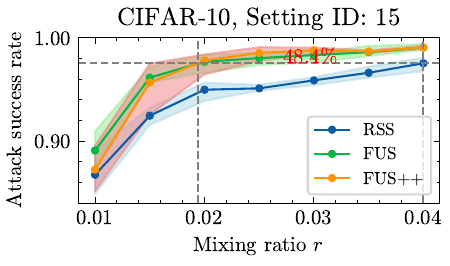}}
\end{center}
\caption{The black-box results of FUS++, FUS, and RSS on CIFAR-10. The \textcolor{red}{red} number is the ratio between the sample volume of FUS++ to RSS, when the FUS++-selected poisoned samples reach the maximum attack success rate of the RSS-selected}
\label{fig:br_c10}
\end{figure*}

\noindent\textbf{White-box Results.} The white-box results on CIFAR-10, CIFAR-100, and ImageNet-10 are shown in \figurename~\ref{fig:wr_c10_c100_i10}. It can be seen that the attack success rate of the poisoned samples selected using FUS and FUS++ is always better than that of the poisoned samples selected using RSS by a large margin for the same mixing ratio. The boosts are about 0.03 to 0.07 for CIFAR-10, 0.04 to 0.1 for CIFAR-100, and 0.03 to 0.05 for ImageNet-10. To compare the data volumes of FUS++ and RSS that reach the same attack strength, we calculate the percentage using linear interpolation, and the results are shown in the red numbers in \figurename~\ref{fig:wr_c10_c100_i10}. FUS++ can save 50\% to 65\% of the data volume to achieve the same attack success rate as RSS. These results indicate that the proposed strategy can improve the efficiency of data poisoning in white-box settings, thereby reducing the number of poisoned samples required. This definitely increases the stealthiness of backdoor attacks.

\noindent\textbf{Black-box Results.} In practice, it is more common that the attacker does not know any prior knowledge about the user's configuration. The black-box results on CIFAR-10, CIFAR-100, and ImageNet-10 are shown in \figurename~\ref{fig:br_c10}, Fig. 20 (Supplementary Materials), and Fig. 21 (Supplementary Materials), respectively. In the majority of black-box settings, poisoned samples selected using FUS and FUS++ consistently have a higher success rate than those selected using RSS at the same mixing ratio. The improvements are about 0.01 to 0.07 for CIFAR-10, 0.02 to 0.1 for CIFAR-100, and 0.02 to 0.07 for ImageNet-10. The only two exceptional cases occur on ImageNet-10 with setting IDs 7 and 15, where the poisoned samples selected using these two selection methods have similar attack success rates. Likewise, we calculate the percentage of poisoned sample size selected by FUS++ to RSS for the same attack intensity. It can be seen that about 20\% to 60\% of the data volume are saved. These results show that the samples selected by FUS and FUS++ have good transferability and can be applied in practice, since it does not require prior knowledge of the DNN architecture, the optimizer, and the training hyperparameters employed by the user. Compared to FUS, our advanced solution, FUS++, consistently outperforms it in ASR across the majority of cases. However, it's important to note that there are still instances where FUS++ faces challenges, particularly when the poison rate is low and the optimizer settings vary between the search and test phases. In our ongoing efforts, we remain dedicated to actively resolving these issues in the future.

\begin{figure}[htbp]
\begin{center}
\includegraphics[width=0.4\textwidth]{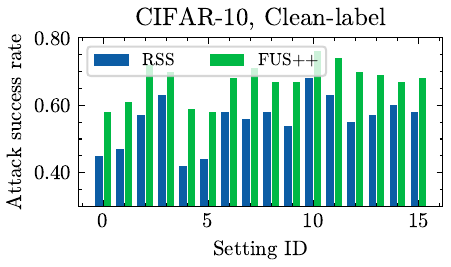}
\end{center}
\caption{The results of FUS++ and RSS on CIFAR-10 with clean-label attack when $r=0.01$.}
\label{fig:cle_lab}
\end{figure}

\noindent\textbf{Results of Clean-label Attack.} The white-box and black-box results exhibited above are under the flip-label attack. Compared to the flip-label attack, the clean-label attack does not yield the inconsistency between the content and label, and is therefore more covert. Here we also evaluate the performance of FUS++ on the clean-label attack, and the results are shown in \figurename~\ref{fig:cle_lab}. It can be seen that when $r=0.01$, the attack success rate using FUS++-selected poisoned samples for clean-label attack is about 0.07 to 0.17 higher than that of using RSS on CIFAR-10.

Together, we validate the effectiveness of FUS++ on different image datasets (CIFAR-10, CIFAR-100, ImageNet-10) and different backdoor attack types (flip-label attack, clean-label attack). The above results demonstrate that the proposed method can significantly improve the poisoning efficiency of backdoor attacks on the image classification task without affecting the clean accuracy. 

\subsection{Comparison with Other State of the arts}
\label{sec:results_sota}

\begin{figure}[htbp]
\begin{center}
\includegraphics[width=0.45\textwidth]{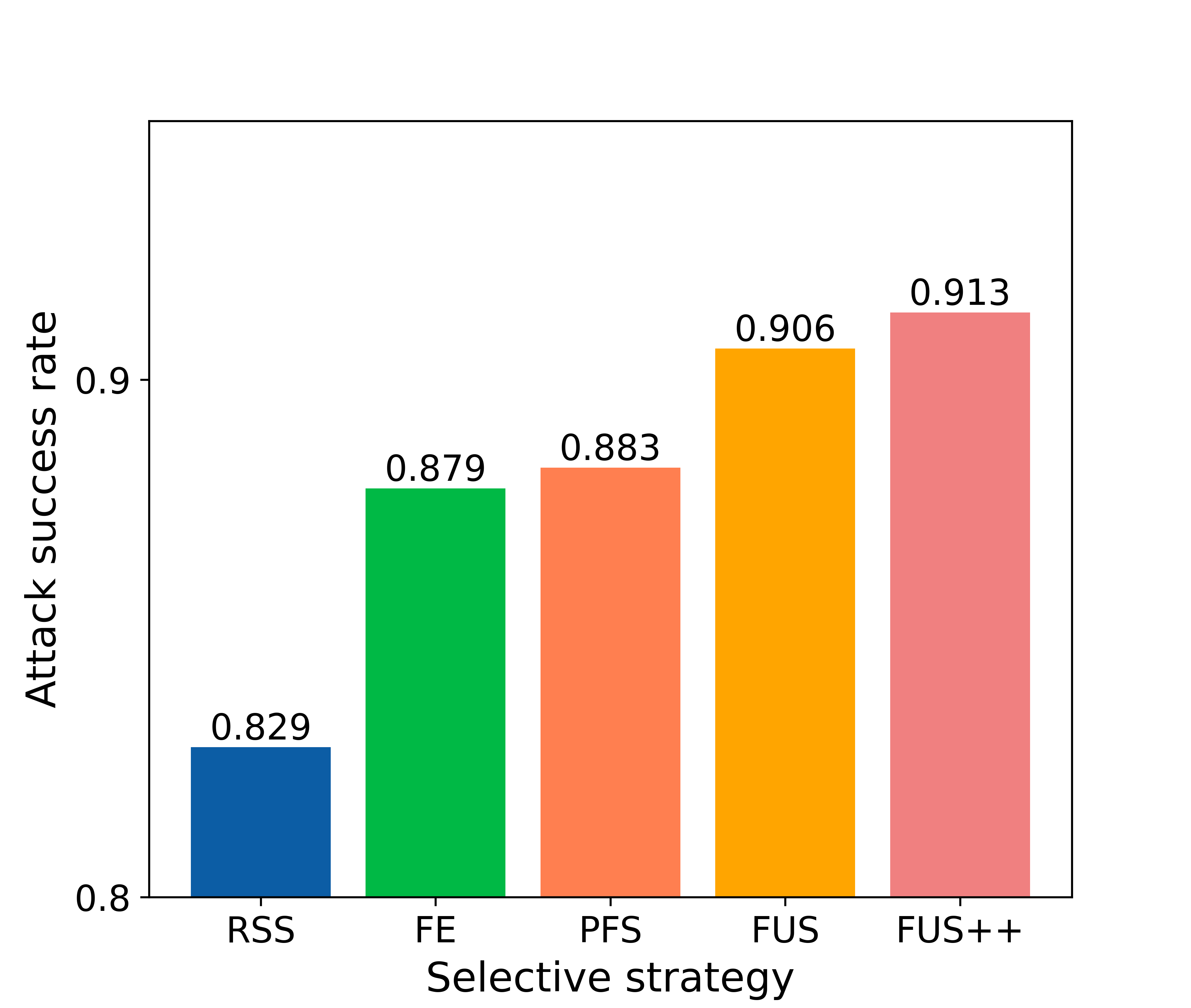}
\end{center}
\caption{The comparison with other state of the arts on CIFAR-10 when $r=0.01$.}
\label{fig:sota}
\end{figure}

As illustrated in \figurename~\ref{fig:sota}, we compare our proposed FUS++ with the baseline (RSS) and other state of the arts (FE \cite{gao2023not}, PFS \cite{li2023proxy}, and FUS \cite{xia2022data}) on CIFAR-10 dataset when $r=0.01$. The results demonstrate that our proposed FUS++ outperform baseline and previous state of the arts.

\subsection{Results on Text Classification}
\label{sec:results_text_classi}
\begin{figure}[htbp]
\begin{center}
\subfigure[]{\includegraphics[width=0.24\textwidth]{\impath/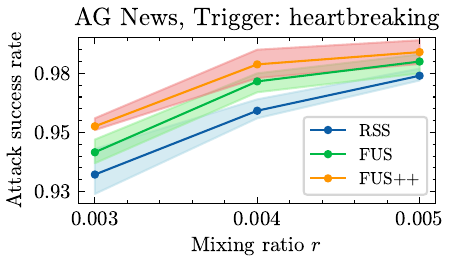}}
\subfigure[]{\includegraphics[width=0.24\textwidth]{\impath/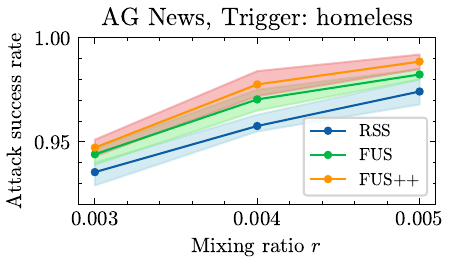}}
\end{center}
\caption{The results of FUS++, FUS, and RSS on AG News.}
\label{fig:ag_new}
\end{figure}

We test FUS and FUS++ here on the text modality. Two words, ``heartbreaking'' and ``homeless'', are set as triggers to poison the AG News training set, individually. The experimental results are shown in \figurename~\ref{fig:ag_new}. The attack success rate using FUS++-selected poisoned samples is about 0.008 to 0.012 higher than that of using RSS on AG News. 

\subsection{Results on Audio Classification}
\label{sec:results_audio_class}
\begin{table}[htbp]
\caption{The results of FUS++, FUS, and RSS on ESC-50.}
\label{tab:esc_50}
\begin{center}
\begin{tabular}{lccc} 
\toprule
$r$  & RSS               & FUS    &FUS++           \\ \midrule
0.06 & 0.770 $\pm$ 0.006 & 0.783 $\pm$ 0.011&0.801$\pm$ 0.007 \\
0.08 & 0.792 $\pm$ 0.011 & 0.809 $\pm$ 0.015&0.817$\pm$ 0.012 \\
\bottomrule
\end{tabular}
\end{center}
\end{table}

We test FUS and FUS++ here on the audio modality. The experimental results are shown in \tablename~\ref{tab:esc_50}. The attack success rate using FUS++-selected poisoned samples is about 0.025 to 0.031 higher than that of using RSS on ESC-50. 

\subsection{Results on Age Regression}
\label{sec:results_age_reg}
\begin{table}[htbp]
\caption{The results of FUS++, FUS, and RSS on Facial Age.}
\label{tab:fac_age}
\begin{center}
\begin{tabular}{lccc} 
\toprule
$r$  & RSS               & FUS     &FUS++          \\ \midrule
0.01 & 3.658 $\pm$ 0.264 & 2.378 $\pm$ 0.088&2.029$\pm$ 0.075 \\
0.02 & 2.282 $\pm$ 0.088 & 1.735 $\pm$ 0.064& 1.513$\pm$ 0.059\\
\bottomrule
\end{tabular}
\end{center}
\end{table}

The above experiments are performed on classification tasks with different input modalities. In this part, we want to evaluate the effectiveness of FUS and FUS++ on the regression task. Unlike the classification task, the attack success rate cannot be used as a metric here; we actually use the RMSE between the output of the infected model and the attack target (i.e., 20 years old) on the poisoned test set as a metric. The smaller the value, the higher the strength of the attack. The experimental results are shown in \tablename~\ref{tab:fac_age}. As can be seen, the improvement of FUS++ for attack intensity is significant.

\subsection{Results on Limited Data}
\label{sec:results_limited}
\begin{table}[htbp]
\caption{The results of FUS++, FUS, and RSS on CIFAR-10 with limited data.}
\label{tab:lim_dat}
\tiny
\begin{center}
\begin{tabular}{cc|cccc|cccc} 
\toprule
\multirow{2}{*}{ID} & \multirow{2}{*}{RSS} & \multicolumn{4}{c|}{FUS}&\multicolumn{4}{c}{FUS++} \\
     &       & All   & 1/2   & 1/3   & 1/4& All   & 1/2   & 1/3   & 1/4   \\ \midrule
 0   & 0.834 & 0.906 & 0.891 & 0.875 & 0.875&0.913&0.895&0.878&0.858 \\
 1   & 0.833 & 0.900 & 0.880 & 0.861 & 0.871&0.899&0.893&0.870&0.840 \\
 2   & 0.890 & 0.947 & 0.938 & 0.919 & 0.931&0.954&0.960&0.946&0.927 \\
 3   & 0.891 & 0.945 & 0.930 & 0.927 & 0.929&0.912&0.921&0.924&0.923 \\
 4   & 0.848 & 0.914 & 0.890 & 0.863 & 0.886&0.905&0.916&0.873&0.865 \\
 5   & 0.831 & 0.904 & 0.877 & 0.864 & 0.873&0.900&0.893 &0.880&0.848\\
 6   & 0.892 & 0.926 & 0.929 & 0.909 & 0.902&0.908&0.918&0.917&0.905 \\
 7   & 0.896 & 0.899 & 0.875 & 0.885 & 0.882&0.891&0.890 &0.883&0.871\\
 8   & 0.864 & 0.935 & 0.909 & 0.901 & 0.888&0.946&0.931 &0.906&0.878\\
 9   & 0.861 & 0.922 & 0.900 & 0.890 & 0.883&0.940&0.920 &0.900&0.879\\
10   & 0.909 & 0.952 & 0.947 & 0.931 & 0.939&0.963&0.964&0.953&0.933 \\
11   & 0.891 & 0.944 & 0.929 & 0.922 & 0.921&0.947 &0.938&0.943&0.923\\
12   & 0.878 & 0.936 & 0.907 & 0.898 & 0.903&0.947&0.931 &0.904&0.873\\
13   & 0.859 & 0.922 & 0.903 & 0.887 & 0.879&0.936&0.919&0.911&0.889 \\
14   & 0.888 & 0.929 & 0.921 & 0.912 & 0.919&0.932&0.943&0.933&0.916 \\
15   & 0.867 & 0.891 & 0.881 & 0.884 & 0.873 &0.873&0.879&0.879&0.876\\
Mean & 0.871 & \textbf{0.923} & 0.907 & 0.895 & 0.897&\textbf{0.923}&0.921&0.906&0.888 \\
\bottomrule
\end{tabular}
\end{center}
\end{table}

The attacker can control all the training data in the above experiments for the classification and regression tasks. However, in some cases, the attacker may only have control over a portion of the training data. Therefore, we conduct experiments in this part on the effectiveness of FUS and FUS++ in this situation. We believe that the impact of the limited data on FUS and FUS++ is mainly focused on the bias of the importance measure. We assume that the attacker has all, 1/2, 1/3, and 1/4 of the training data on CIFAR-10, respectively, and the experimental results are shown in \tablename~\ref{tab:lim_dat}. It can be seen that the boost from FUS and FUS++ is gradually decreasing as the proportion of attacker-controlled training data decreases; however, even in the case of 1/4, the FUS++-selected poisoned samples still have a higher attack success rate than the RSS-selected poisoned samples.

\subsection{Attribution Studies}
\label{sec:att_studies}
In this part, we want to figure out what makes the poisoned samples selected by FUS more efficient than those selected by RSS. We conduct experiments on CIFAR-10.

\begin{figure*}[htbp]
\begin{center}
\subfigure[]{\includegraphics[width=0.19\textwidth]{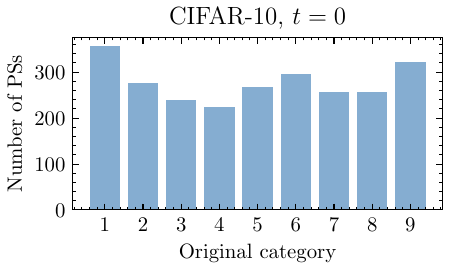}}
\subfigure[]{\includegraphics[width=0.19\textwidth]{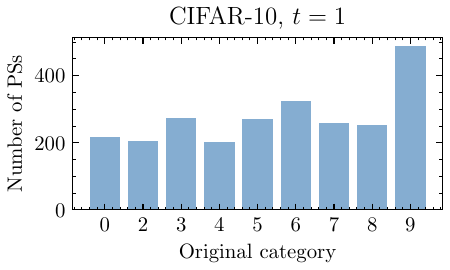}}
\subfigure[]{\includegraphics[width=0.19\textwidth]{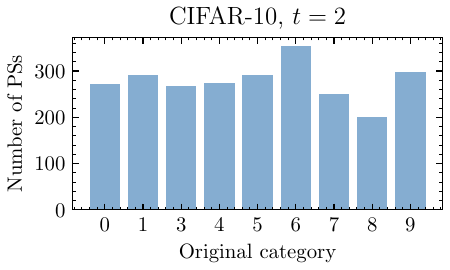}}
\subfigure[]{\includegraphics[width=0.19\textwidth]{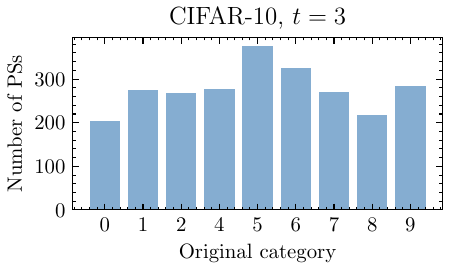}}
\subfigure[]{\includegraphics[width=0.19\textwidth]{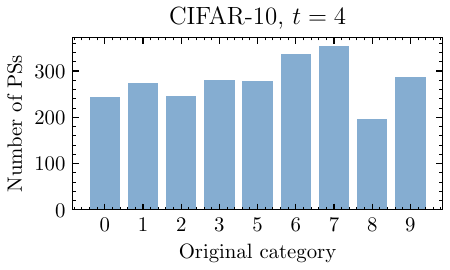}} \\
\subfigure[]{\includegraphics[width=0.19\textwidth]{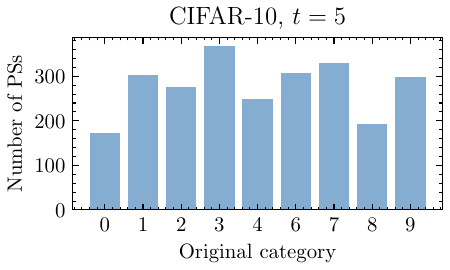}}
\subfigure[]{\includegraphics[width=0.19\textwidth]{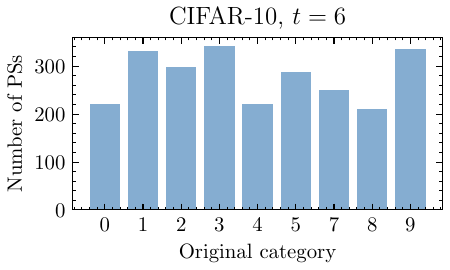}}
\subfigure[]{\includegraphics[width=0.19\textwidth]{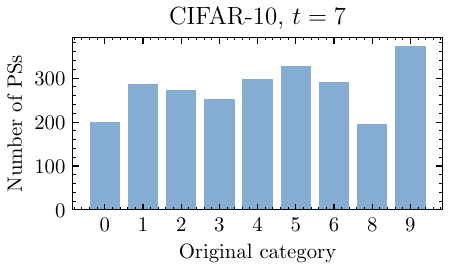}}
\subfigure[]{\includegraphics[width=0.19\textwidth]{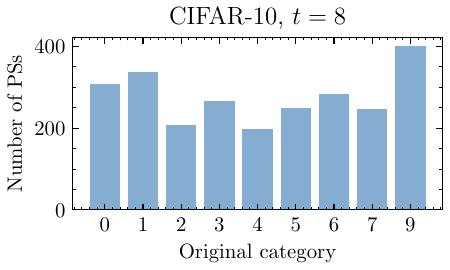}}
\subfigure[]{\includegraphics[width=0.19\textwidth]{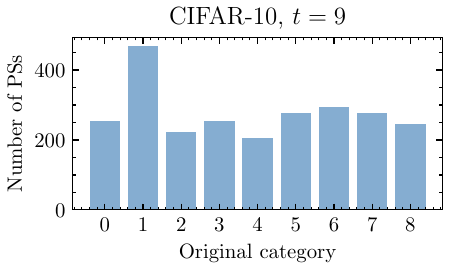}}
\end{center}
\caption{The original category statistics of the FUS-selected poisoned samples for different $t$ on CIFAR-10 when $r=0.01$. We run it 5 times independently, so there are $5*r*50000=2500$ samples for each count. The 10 categories in CIFAR-10 are 0: ``airplane'', 1: ``automobile'', 2: ``bird'', 3: ``cat'', 4: ``deer'', 5: ``dog'', 6: ``frog'', 7: ``horse'', 8: ``ship'', and 9: ``truck''.}
\label{fig:dis_c10}
\end{figure*}

\noindent\textbf{Class Distribution of the Poisoned Samples.}
The first possible reason we consider is that FUS may select more samples from categories that are more relevant to the attack target $t$. Therefore, we count the original categories for the FUS-selected poisoned samples with different $t$. The results are shown in \figurename~\ref{fig:dis_c10}, and the most categories are shown in \tablename~ \ref{tab:mos_cat}. These results seem to support our assumption. For example, ``automobile'' and ``truck'' are mutually the most original categories of the FUS-selected poisoned samples, which is also consistent with human intuition.

\begin{table}[htbp]
\caption{The most original category in the FUS-selected poisoned samples and its percentage on CIFAR-10. It can be seen that there is a correlation between $t$ and the most category.}
\label{tab:mos_cat}
\begin{center}
\begin{tabular}{llc} 
\toprule
Attack target $t$ & Most selected category & Percentage \\ \midrule
0: ``airplane''   & 1: ``automobile''      & 14.3\%     \\
1: ``automobile'' & 9: ``truck''           & 19.6\%     \\
2: ``bird''       & 6: ``frog''            & 14.2\%     \\
3: ``cat''        & 5: ``dog''             & 15.1\%     \\
4: ``deer''       & 7: ``horse''           & 14.2\%     \\
5: ``dog''        & 3: ``cat''             & 14.8\%     \\
6: ``frog''       & 3: ``cat''             & 13.7\%     \\
7: ``horse''      & 9: ``truck''           & 15.0\%     \\
8: ``ship''       & 9: ``truck''           & 16.1\%     \\
9: ``truck''      & 1: ``automobile''      & 18.8\%     \\
\bottomrule
\end{tabular}
\end{center}
\end{table}

Naturally, the next question is whether samples \textit{randomly} sampled according to the same class distribution as the poisoned samples selected by FUS will also yield the same attack performance. We verify this problem and the results are shown in \tablename~ \ref{tab:rss_u_s}. In both distributions, the attack success rates of the poisoned samples selected using RSS are similar, and both are far from the FUS-selected samples. This suggests that the fundamental reason why the FUS-selected poisoned samples work well is not because of the class distribution, but the samples themselves. \textbf{The class distribution is a symptom, not a cause}.

\begin{table}[htbp]
\caption{The attack success rate of RSS on CIFAR-10 and VGG-13 with the Uniform distribution (U) and with the same class distribution (S) as the FUS-selected samples.}
\label{tab:rss_u_s}
\begin{center}
\begin{tabular}{lccc} 
\toprule
Attack target $t$ & RSS with U        & RSS with S        & FUS               \\ \midrule
0: ``airplane''   & 0.831 $\pm$ 0.018 & 0.836 $\pm$ 0.009 & 0.910 $\pm$ 0.008 \\
1: ``automobile'' & 0.834 $\pm$ 0.016 & 0.839 $\pm$ 0.007 & 0.922 $\pm$ 0.016 \\
2: ``bird''       & 0.828 $\pm$ 0.021 & 0.841 $\pm$ 0.002 & 0.887 $\pm$ 0.014 \\
3: ``cat''        & 0.820 $\pm$ 0.018 & 0.816 $\pm$ 0.008 & 0.875 $\pm$ 0.009 \\
4: ``deer''       & 0.826 $\pm$ 0.017 & 0.823 $\pm$ 0.007 & 0.882 $\pm$ 0.009 \\
5: ``dog''        & 0.838 $\pm$ 0.005 & 0.836 $\pm$ 0.007 & 0.905 $\pm$ 0.004 \\
6: ``frog''       & 0.840 $\pm$ 0.019 & 0.826 $\pm$ 0.011 & 0.907 $\pm$ 0.008 \\
7: ``horse''      & 0.835 $\pm$ 0.008 & 0.847 $\pm$ 0.012 & 0.919 $\pm$ 0.004 \\
8: ``ship''       & 0.819 $\pm$ 0.005 & 0.829 $\pm$ 0.006 & 0.887 $\pm$ 0.012 \\
9: ``truck''      & 0.840 $\pm$ 0.007 & 0.837 $\pm$ 0.009 & 0.908 $\pm$ 0.018 \\
\bottomrule
\end{tabular}
\end{center}
\end{table}

\begin{figure*}[htbp]
\begin{center}
\subfigure[]{\includegraphics[width=0.32\textwidth]{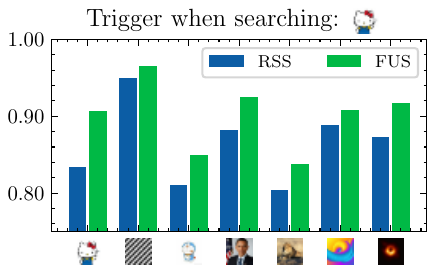}\label{fig:trigger_a}}
\subfigure[]{\includegraphics[width=0.32\textwidth]{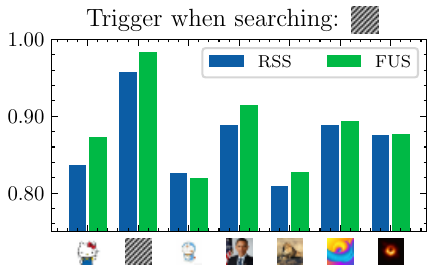}}
\subfigure[]{\includegraphics[width=0.32\textwidth]{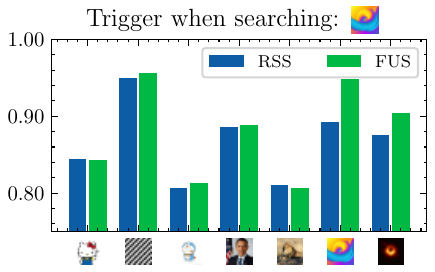} \label{fig:trigger_c}} 
\end{center}
\caption{The attack success rate when the trigger used for attack differs from the trigger used for search. It can be seen that the requirements for important poisoned samples are not exactly the same between different triggers.}
\label{fig:trigger}
\end{figure*}

\noindent\textbf{Trigger of the Poisoned Attacks.}
Furthermore, we would like to know if the selection of these samples is related to the trigger form. This question determines whether some features of the original clean samples can help with the backdoor injection, or whether it is caused by the cooperation of these samples and the trigger. We verify this question by making the trigger used in the FUS search different from the trigger used in the poisoning, and the results are shown in \figurename~\ref{fig:trigger}. It can be seen from the results that the FUS-selected poisoned samples and the trigger are correlated to some extent. For example, the samples selected using the ``hello kitty'' image as the trigger are also suited to other triggers shown in \figurename~\ref{fig:trigger_a}, while the samples selected using the ``rainbow-colored tube'' image as the trigger are not well suited to other triggers shown in \figurename~\ref{fig:trigger_c}. The transfer performance of FUS-selected samples between different triggers should be related to the similarity between these triggers. Therefore, \textbf{the high efficiency of the FUS-selected samples is not just a property of the sample itself, it is the result of the interaction of the sample and the trigger}.

\begin{figure}[t!]
    \setlength{\belowcaptionskip}{-0.5cm}
	\centering
	\includegraphics[width=0.9\linewidth]{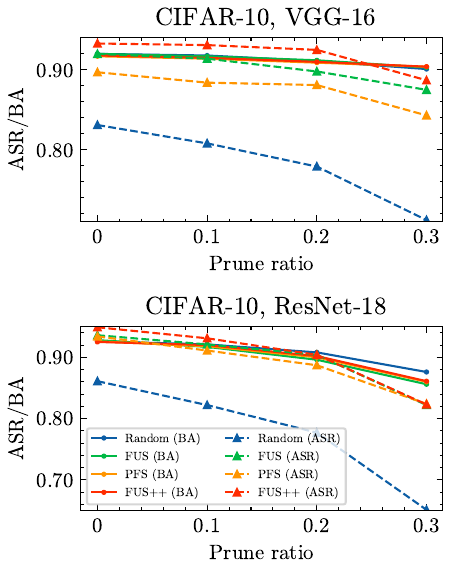}
	\caption{Defense results of pruning, where the number of clean samples owned by the defender is 50 and the trigger is Blended. }
	\label{fig:deference}
\end{figure}

\noindent\textbf{Against Defence Methods.}
Finally, we analyze the resilience of the RSS, PFS \cite{li2023proxy}, FUS and FUS++ against Pruning-based backdoor defense \cite{liu2018fine}. As depicted in Fig. \ref{fig:deference}, FUS and FUS++ emerge as a robust defense against Pruning \cite{liu2018fine} in contrast to Random selection (RSS) and PFS \cite{li2023proxy}.

\section{Conclusion} \label{sec:con}
This study demonstrates that the poisoning efficiency of backdoor attacks can be substantially improved by selecting suitable poisoned samples. We propose a novel selection strategy named FUS++ to iteratively filter and update a poisoned sample pool to achieve this goal. Specifically, we adopt the forgetting events of the samples to indicate the contribution of different poisoned samples and use the curvature of the loss surface to analyses the effectiveness of this phenomenon. Accordingly, we combine forgetting events and curvature of different samples to conduct a sample yet efficient sample selection strategy. The experimental results on image classification (CIFAR-10, CIFAR-100, ImageNet-10), text classification (AG News), audio classification (ESC-50), and age regression (Facial Age) consistently demonstrate the effectiveness of our method.

\section*{Acknowledgments}
The work was supported in part by the National Natural Science Foundation of China under Grands U19B2044 and 61836011.

\bibliographystyle{ieee_fullname}
\bibliography{./references.bib}

\begin{IEEEbiography}[{\includegraphics[width=1in,height=1.25in,clip,keepaspectratio]{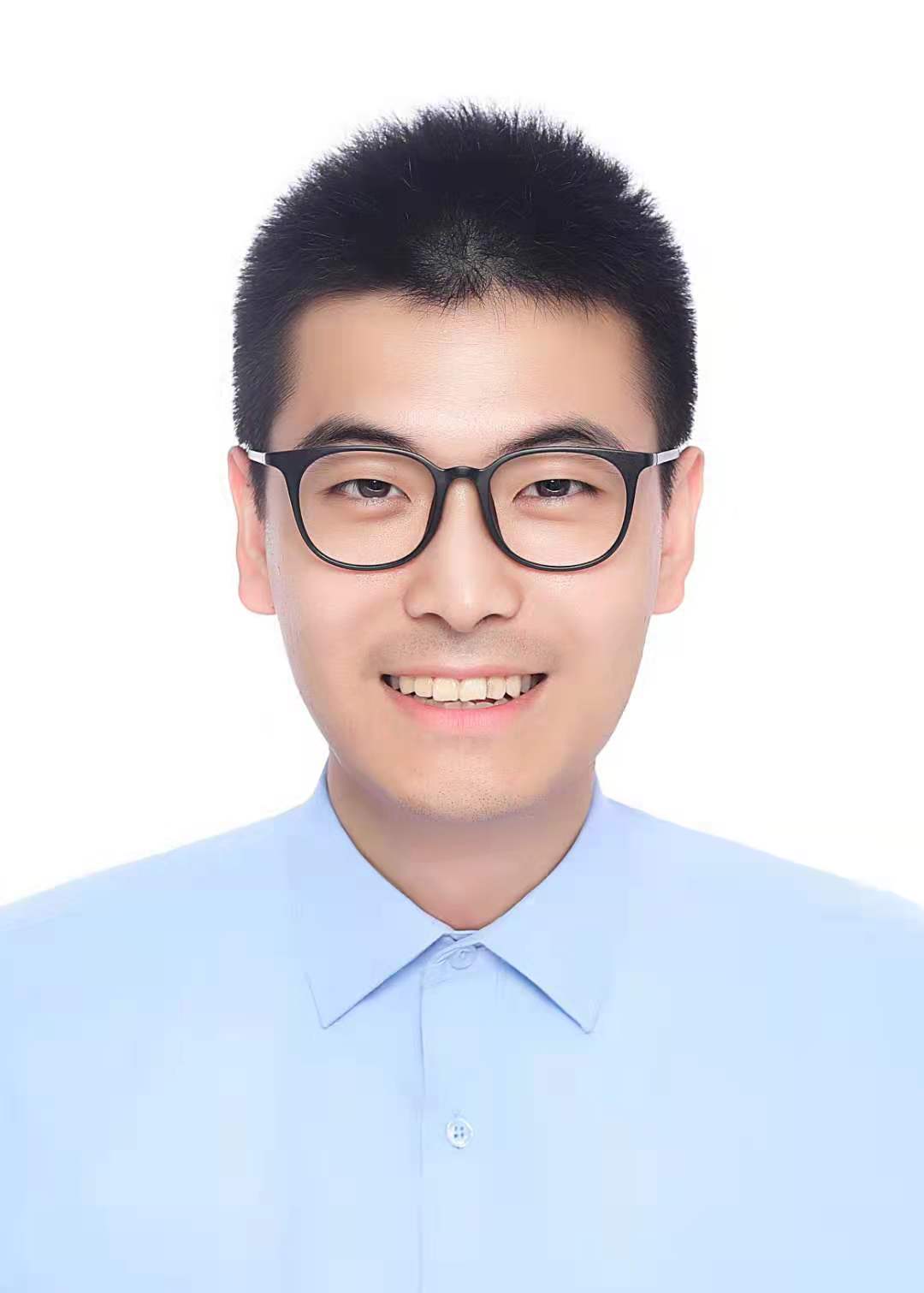}}]{Ziqiang Li} received the B.E. degree in electronic science and technology from University of Science and Technology of China (USTC), Hefei, China, in 2019 and is pursuing the Ph.D. degree in Information and Communication Engineering from University of Science and Technology of China (USTC), Hefei, China. His research interests include backdoor learning, deep generative models, and machine learning.
\end{IEEEbiography}

\begin{IEEEbiography}[{\includegraphics[width=1in,height=1.25in,clip,keepaspectratio]{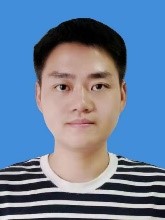}}]{Pengfei Xia} received the B.E. degree from China University of Mining and Technology (CUMT), Xuzhou, China, in 2015 and is pursuing the Ph.D. degree from University of Science and Technology of China (USTC), Hefei, China. His research interests include adversarial examples, backdoor learning, and secure deep learning.
\end{IEEEbiography}

\begin{IEEEbiography}[{\includegraphics[width=1in,height=1.25in,clip,keepaspectratio]{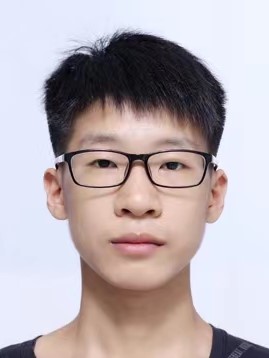}}]{Hong Sun} received the B.E. degree in electronic and information engineering from Dalian University of Technology (DUT), Dalian, China, in 2021 and is pursuing the Masterdegree in Electronics and Communication Engineering from University of Science and Technology of China (USTC), Hefei, China. His research interests include backdoor learning, generative adversarial networks, and machine learning.
\end{IEEEbiography}

\begin{IEEEbiography}[{\includegraphics[width=1in,height=1.25in,clip,keepaspectratio]{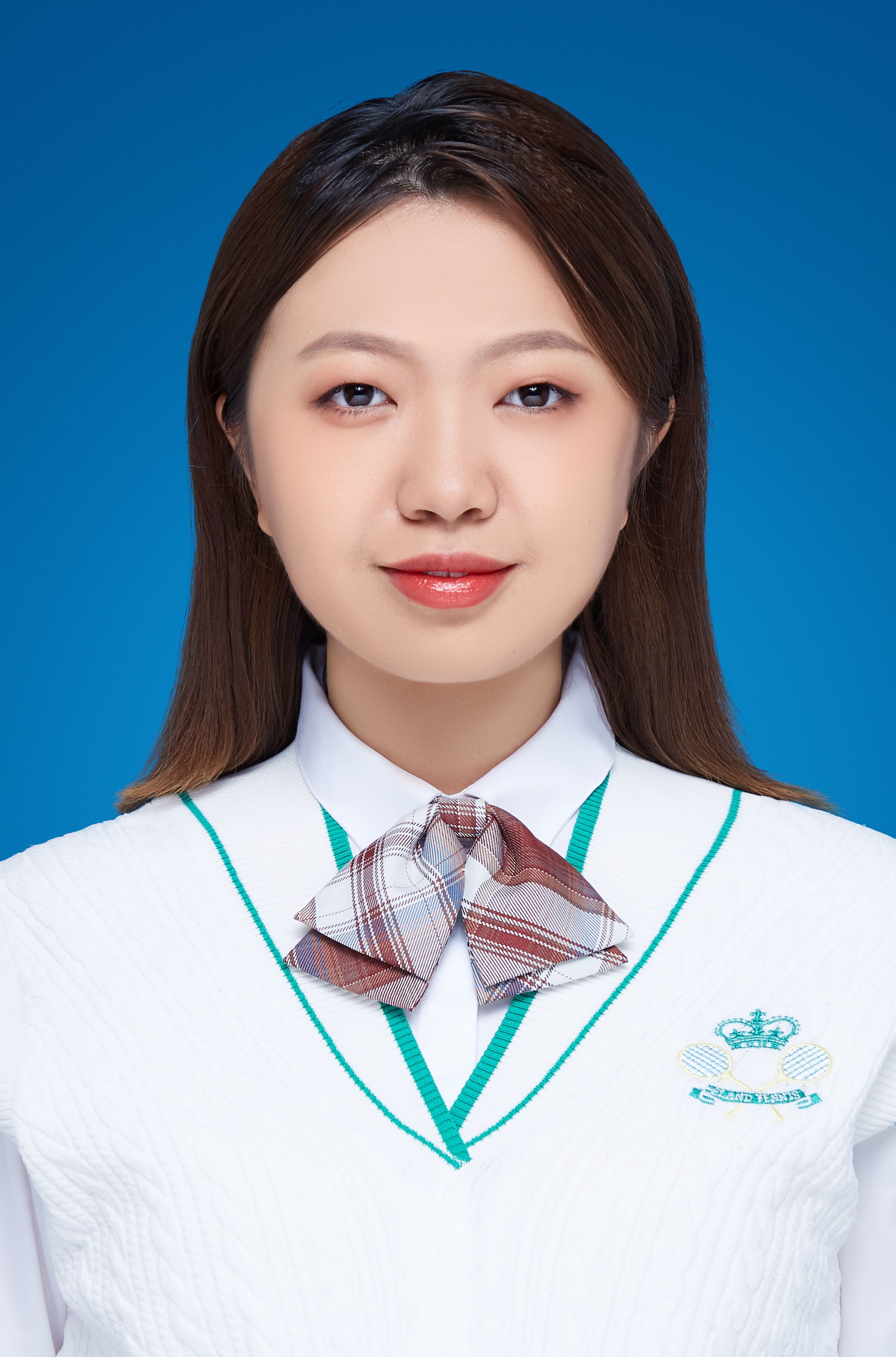}}]{Yueqi Zeng} received the B.E. degree in Electronic and Information Engineering from Northwestern Polytechnical University (NPU), Xi'an, China, in 2021 and is pursuing the Master degree in Electronic Information from University of Science and Technology of China(USTC), Hefei, China. Her current research interests concentrate on AI security, especially backdoor attacks.
\end{IEEEbiography}

\begin{IEEEbiography}[{\includegraphics[width=1in,height=1.25in,clip,keepaspectratio]{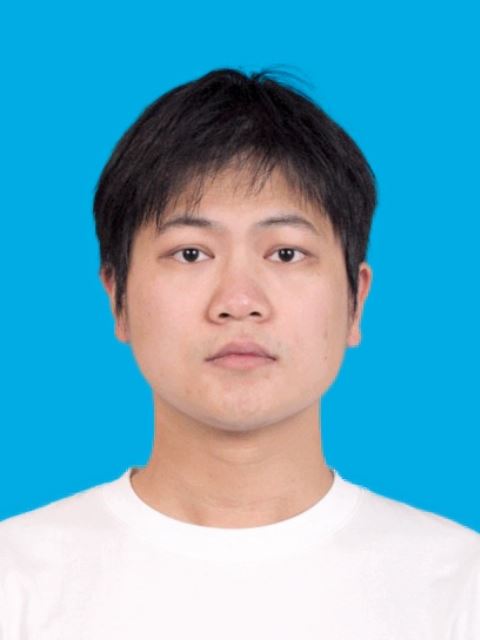}}]{Wei Zhang} received the B.E. degree in Automation from University of Science and Technology of China (USTC), Hefei, China, in 2017 and is pursuing the Ph.D. degree in Information and Communication Engineering from University of Science and Technology of China (USTC), Hefei, China. His research interests include machine learning, human-computer interaction and mixed reality.
\end{IEEEbiography}

\begin{IEEEbiography}[{\includegraphics[width=1in,height=1.25in,clip,keepaspectratio]{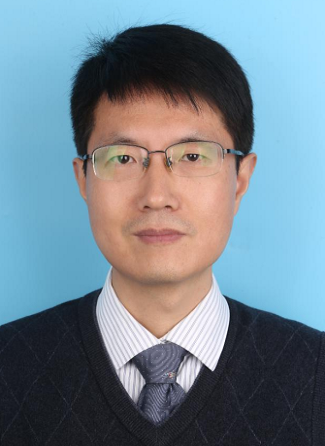}}]{Bin Li} received the B.S. degree from the Hefei University of Technology (HFUT), China, in 1992, the M.S. degree from the Institute of Plasma Physics, Chinese Academy of Sciences, Hefei, China, in 1995, and the Ph.D degree from the University of Science and Technology of China (USTC), Hefei, China, in 2001. He is currently a Professor with the School of Information Science and Technology, USTC. He has authored or co-authored over 40 refereed publications. His current research interests include evolutionary computation, pattern recognition, and human-computer interaction. Dr. Li is the Founding Chair of the IEEE Computational Intelligence Society Hefei Chapter, a Counselor of the IEEE USTC Student Branch, a Senior Member of the Chinese Institute of Electronics (CIE), and a member of the Technical Committee of the Electronic Circuits and Systems Section of CIE.
\end{IEEEbiography}
\begin{figure*}[htbp]
\begin{center}
\subfigure[]{\includegraphics[width=0.19\textwidth]{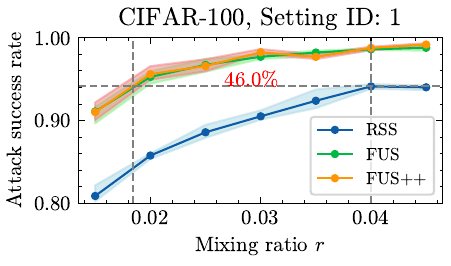}}
\subfigure[]{\includegraphics[width=0.19\textwidth]{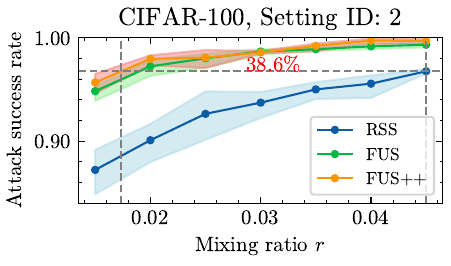}}
\subfigure[]{\includegraphics[width=0.19\textwidth]{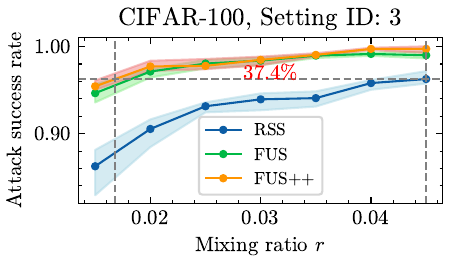}}
\subfigure[]{\includegraphics[width=0.19\textwidth]{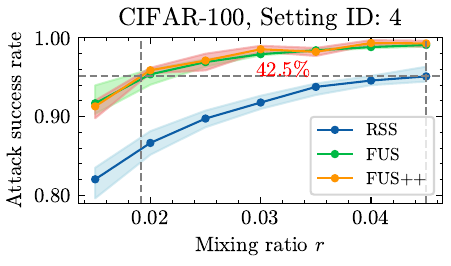}}
\subfigure[]{\includegraphics[width=0.19\textwidth]{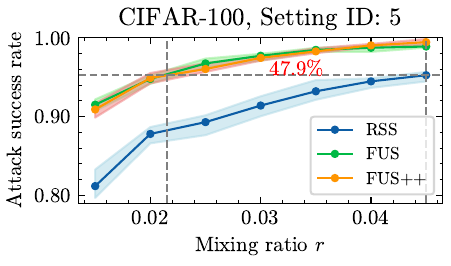}} \\
\subfigure[]{\includegraphics[width=0.19\textwidth]{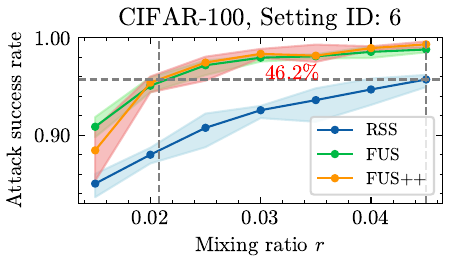}}
\subfigure[]{\includegraphics[width=0.19\textwidth]{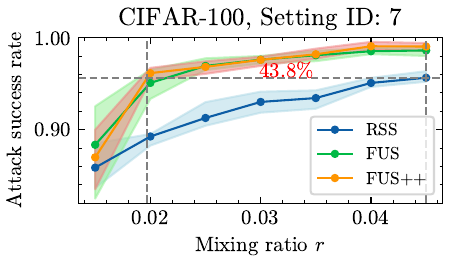}}
\subfigure[]{\includegraphics[width=0.19\textwidth]{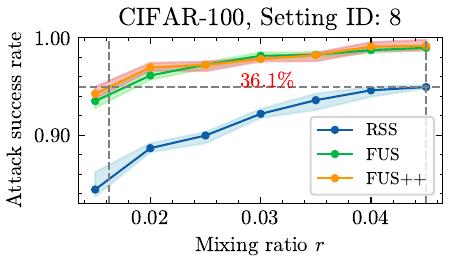}}
\subfigure[]{\includegraphics[width=0.19\textwidth]{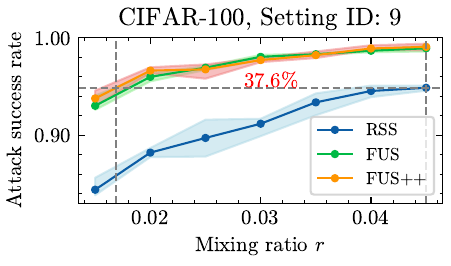}}
\subfigure[]{\includegraphics[width=0.19\textwidth]{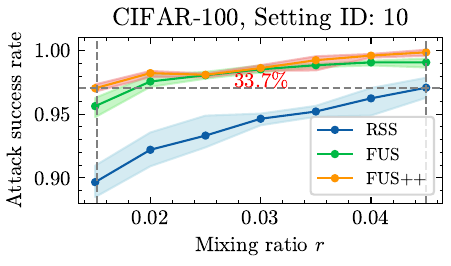}} \\
\subfigure[]{\includegraphics[width=0.19\textwidth]{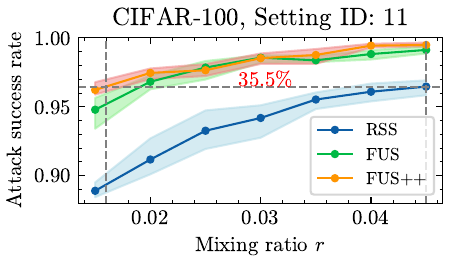}}
\subfigure[]{\includegraphics[width=0.19\textwidth]{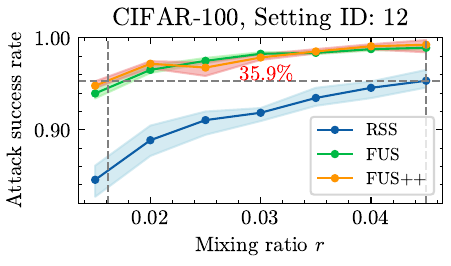}}
\subfigure[]{\includegraphics[width=0.19\textwidth]{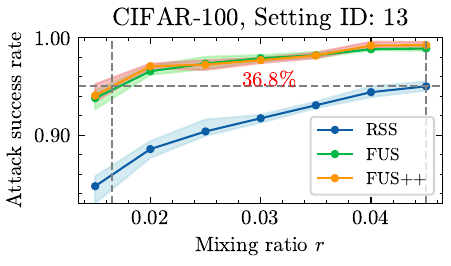}}
\subfigure[]{\includegraphics[width=0.19\textwidth]{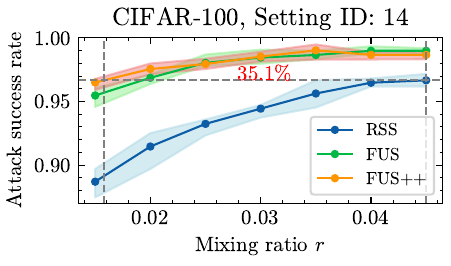}}
\subfigure[]{\includegraphics[width=0.19\textwidth]{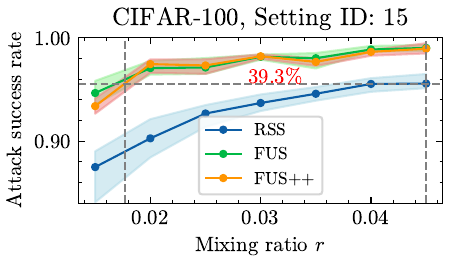}}
\end{center}
\caption{The black-box results of FUS++, FUS and RSS on CIFAR-100. The \textcolor{red}{red} number is the ratio between the sample volume of FUS++ to RSS, when the FUS++-selected poisoned samples reach the maximum attack success rate of the RSS-selected}
\label{fig:br_c100}
\end{figure*}

\begin{figure*}[htbp]
\begin{center}
\subfigure[]{\includegraphics[width=0.19\textwidth]{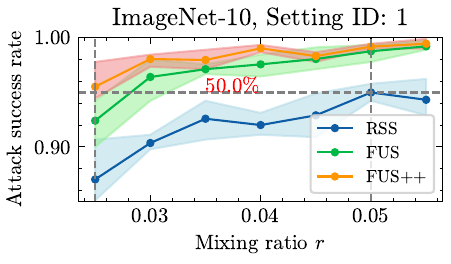}}
\subfigure[]{\includegraphics[width=0.19\textwidth]{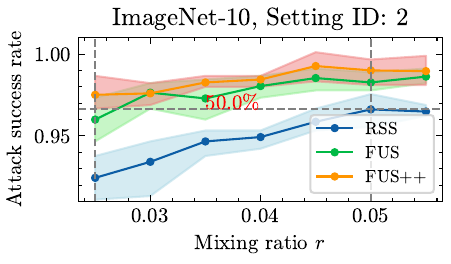}}
\subfigure[]{\includegraphics[width=0.19\textwidth]{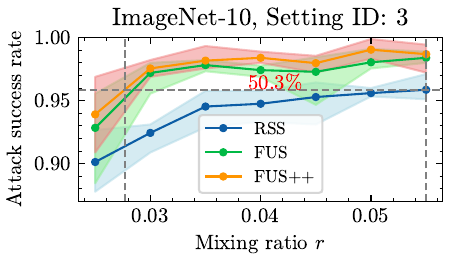}}
\subfigure[]{\includegraphics[width=0.19\textwidth]{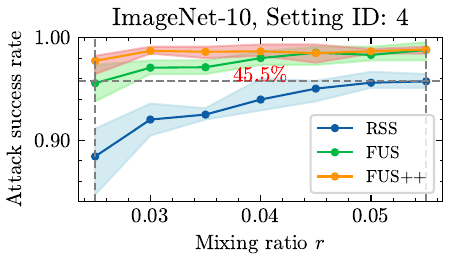}}
\subfigure[]{\includegraphics[width=0.19\textwidth]{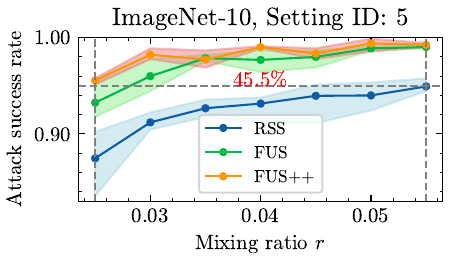}} \\
\subfigure[]{\includegraphics[width=0.19\textwidth]{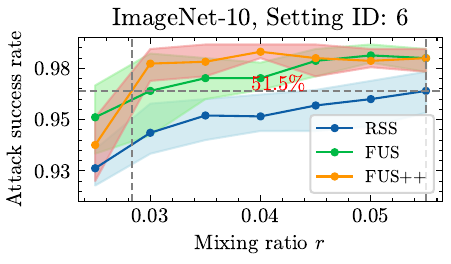}}
\subfigure[]{\includegraphics[width=0.19\textwidth]{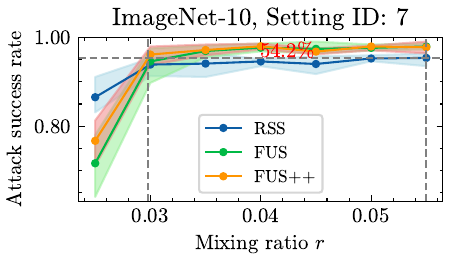}}
\subfigure[]{\includegraphics[width=0.19\textwidth]{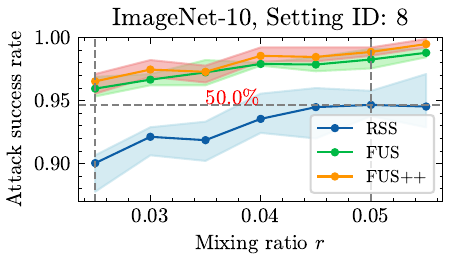}}
\subfigure[]{\includegraphics[width=0.19\textwidth]{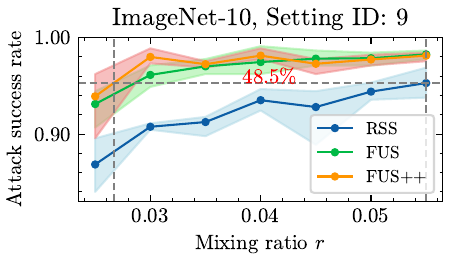}}
\subfigure[]{\includegraphics[width=0.19\textwidth]{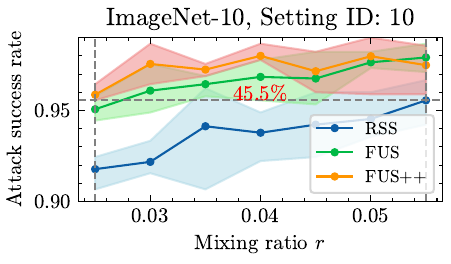}} \\
\subfigure[]{\includegraphics[width=0.19\textwidth]{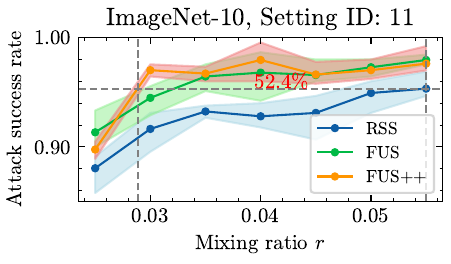}}
\subfigure[]{\includegraphics[width=0.19\textwidth]{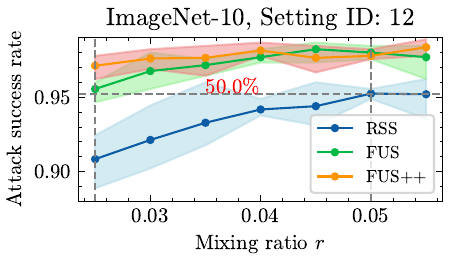}}
\subfigure[]{\includegraphics[width=0.19\textwidth]{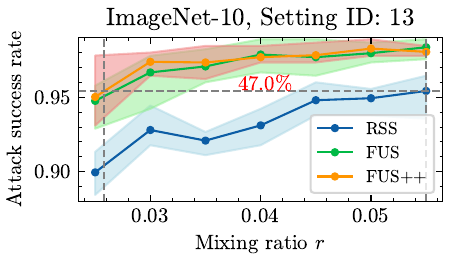}}
\subfigure[]{\includegraphics[width=0.19\textwidth]{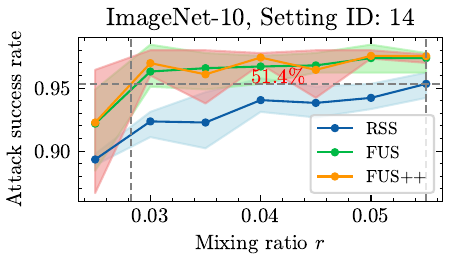}}
\subfigure[]{\includegraphics[width=0.19\textwidth]{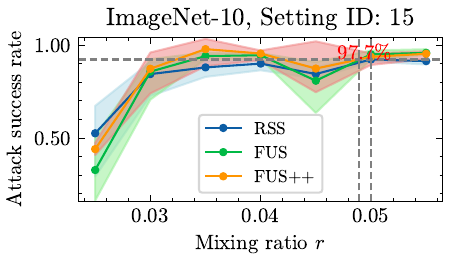}}
\end{center}
\caption{The black-box results of FUS++, FUS and RSS on ImageNet-10. The \textcolor{red}{red} number is the ratio between the sample volume of FUS++ to RSS, when the FUS++-selected poisoned samples reach the maximum attack success rate of the RSS-selected}
\label{fig:br_i10}
\end{figure*}
\end{document}